\documentstyle[aas2pp4]{article}
\begin{document}

\title{Review: Accretion disk theory}

\author{Matias Montesinos Armijo}
%\author{Revised by Jeannette Barnes, May 1995}
\author{\it Departamento de Astronom\'ia y Astrof\'isica, Pontificia Universidad Cat\'olica de Chile, Santiago de Chile}
%\authoraddr{}

\begin{abstract}

In this paper I review and discuss the basic concepts of accretion disks,  focused especially on the case of accretion disks around black holes. 
The well known $\alpha$-model is revisited, showing the strengths and weaknesses
of the model. Other turbulent viscosity prescription, based on the Reynolds number, that may improve our understanding of the accretion paradigm is  discussed. 
A simple but efficient  mathematical model of a self-gravitating accretion disk, as well as observational evidence of these objects, are also included.

\end{abstract}

\twocolumn

\section{Brief review of accretion disk theory}

Accretion disks are expected to be associated to different astrophysical objects in the universe
and, among them it is possible to mention; disks around young stars (T Tauri), probably sites
of planetary formation; disks in close binary systems like $\beta$-Lyra, cataclysmic variables,
novae, X-ray binaries, etc., and AGNs in general.

The basic idea of accretion disks was proposed soon after the confirmation of quasars, and its classical picture established
at the beginning of the 1970s.

When the gas infalls into the compact object, some of the binding gravitational energy of the
gas is released due the viscosity present in the accretion flow. 
The  amount of gravitational energy liberated during the process is huge. This was first pointed out by
Zeldovich (1964), Zeldovich \& Novikov  (1967), and independently by Salpeter   (1964). However, they 
considered an isolated compact object converting gravitational energy into radiation during its collapse.

Consequently to these historical papers,  several models explaining observed energetic phenomena like
X-ray binaries appeared. For instance Shklovsky  (1967)   explained X-ray
emissions from Sco X-1\footnote{Sco X-1 was the first extrasolar X-ray source, discovered in 1962 
(Giacconi et al. )} as accretion of matter from a companion onto a neutron star.
However at the time, the majority of investigators 
assumed instead Bremsstrahlung radiation of an optically thin layer of hot plasma to explain energetic objects.

Under the same argumentation that gravitational energy can be converted into radiation by viscous forces,
Lynden-Bell proposed in 1969  that the energy source of quasars can be 
explained by viscous accretion disks.

The accretion disk model grew in popularity and began to be recognized as the theory capable to describe important 
energetic emissions observed in the Universe. It was in 1973 that Shakura \& Sunyaev  proposed 
a fundamental theory of accretion disks known as the \textit{standard accretion disk model} or simply the
\textit{$\alpha$-disk model}. Very soon, in the same year, it was generalized to the relativistic version by Novikov \& Thorne (1973).

The Shakura \& Sunyaev model, described a relatively cold disk (in range $\sim 10^2 - 10^5~\rm K$ for typical case of quasars)
due an efficient process of conversion and evacuation of gravitational
energy into radiation. The disk was also assumed \textit{geometrically thin} and \textit{optically thick}. They introduced a clever
parameterization for the turbulent viscosity, needed to guarantee the transfer of angular momentum and the process
of accretion flow. 

Around the 80s different features of accretion disk were explored such as hot disks, radiatively inefficient disk, 
geometrically thick disks, etc.

It is the case, for instance, of the \textit{Polish doughnut} disk, consisting in an optically thick torus
introduced by Paczy\'nski \& Wiita in 1980,  and later, when advection were recognized as an
important ingredient, the \textit{slim} disk appeared (Abramowicz  1988),  giving a new branch of
accretion disk, where the advective energy transport was included in the energy balance, producing super-Eddington accretion rates.

The importance of these advective models, called ADAFs (Advection-Dominated-Accretion-Flows),
grew up in the 1990s, and were developed by several authors 
(e.g., Narayan \& Yi  (1994), Abramowicz \& Lasota  (1995)).

All these models showed a good agreement with observations. Accretion disk is now the accepted paradigm 
in order to describe several energetic phenomena either on small scales, like disks around Young stellar object (YSO), or 
Cataclysmic variable stars (CV) or on large scales, like AGNs or quasars.
Table (\ref{AstroObjects}), summarizes some characteristic properties,
such as the mass, size, temperature, and luminosity of the accretion disk supposed to be at the origin 
of the indicated astrophysical object.

%%%%%%%%%%%%%%%%%%%%%%%%%%%%%%%%%%table 1%%%%%%%%%%%%%%%%%%%%%%
\begin{table}[hu] \small
\caption{Properties of the accretion disk in various astrophysical objects.}            
\centering                         
\begin{tabular}{c c c c c c} 
\\      
\hline\hline                 
Central object &Central object mass & Disk size &  Disk mass  & Temperature       & Luminosity\\   
\hline                        
CV/SSXS        & $\sim M_\odot$        &  $\sim R_\odot$ & $ \ll   M_\odot$ &  $\sim 10^4 - 10^6$ K &  $\sim 10^0 - 10^2 L_\odot$ \\     

XB/BHB        & $\gtrsim 3M_\odot$        &  $\sim R_\odot$ &  $ \ll   M_\odot$   &  $\sim 10^4 - 10^9$ K &  $\sim 10^0 - 10^5 L_\odot$ \\     

YSO        & $\sim M_\odot$        &  $\sim 100$ AU &  $\sim M_\odot$   &   $\sim 10^1 - 10^4$ K &  $\sim L_\odot$ \\     

AGN        & $\sim 10^5-10^9 M_\odot$        &  $\sim$pc &  $\sim 10^5-10^9 M_\odot$    &  $\sim 10^3 - 10^6$ K &  $\sim 10^{10} - 10^{13} L_\odot$ \\     
\hline
\end{tabular}
\newline 
\newline
The used nomenclature reads- CV: cataclysmic variable, SSXS: supersoft X-ray source, XB: X-ray binary,
BHB: black hole binary, YSO: young stellar object, AGN: active galactic nuclei.
\label{AstroObjects}
\end{table}
%%%%%%%%%%%%%%%%%%%%%%%%%%%%%%%%%%%%%%%%%%%%%%%%%%%%%%%%%%%%%%%%%%%%%%%

%%%%%%%%%%%%%%%%%%%%%%%%%%%%%%%%%%%%%%%%%% fig 1 %%%%%%%%%%%%%%%%%%%%%%%%%%%%%%%%%%%%%%%%%%%%%%%%%
\begin{figure}[h]
\centering
\plotone{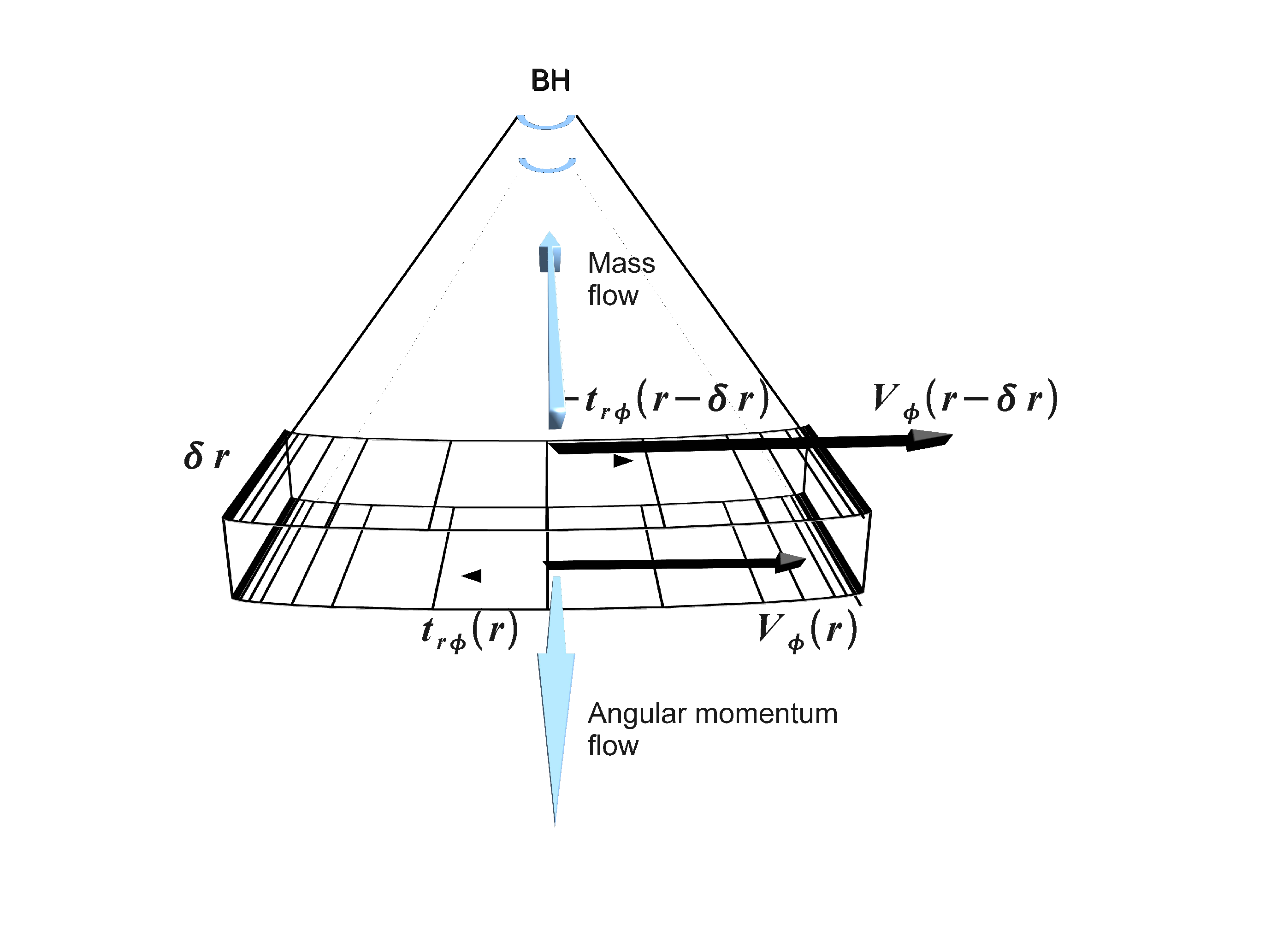}
\caption{Schematic diagram of an accretion disk process. 
The inner edge  of the annulus at $(r-\delta r)$  receives a force $-t_{r \varphi}(r-\delta r)$ per unit surface, gaining angular momentum. Likewise, the outer 
edge of the annulus at $r$ receives a force $t_{r \varphi}$ per unit surface, giving away  angular momentum. Consequently, angular momentum is transfered outward while 
mass is accreted inward.}
\label{PictoDisk}
\end{figure}
%%%%%%%%%%%%%%%%%%%%%%%%%%%%%%%%%%%%%%%%%%%%%%%%%%%%%%%%%%%%%%%%%%%%%%%%%%%%%%%%%%%%%%%%%%%%%%%%%%

Summarizing, the central idea of accretion disks lies in the fact that the accreting masses possess a
considerable amount of angular momentum per mass unit  which has to be removed in order to be accreted into the central object. 
What causes the lost of angular momentum is the friction caused by turbulent viscosity working between adjacent gas 
layers in the disk. The faster inner layer loses angular momentum and infalls slightly, while the next (slower) outer layer
gains angular momentum, which is given away to the next outer layer, and so on, resulting in a continuous flow
toward the centre  whereas 
angular momentum is transported to the outer region (see Figure \ref{PictoDisk} for a pictorial view
of the accretion process). If the disk is massive enough in the external region,  
the gain of angular momentum (of the very last outer layer)  will cause an expansion of the disk
because there is no other shell of gas to give up the gained extra momenta.
In that case, material in that layer is transferred outward.

The friction of the disk heats up the gas, resulting in a continuous radiation emission, which is
believed to be the source of AGNs, quasars, X-ray binaries, luminosities (among other phenomena) as mentioned above.

%The key ingredient are; the viscosity prescription, the self-gravity of the 
%disk, and the dependence of time of all processes.

\section{Basic equations} \label{BasicEquations}

In order to obtain a realistic model of an accretion disk we need to have an adequate
description of the flow; that is why the complete set of hydrodynamic equations must be
solved.

Fluid motion can be completely determined by conservation laws. They state that at all times, a certain number of properties, e.g., mass,
generalized momentum and energy, are conserved quantities.

We start the analysis with the Navier-Stokes equations, describing
the motion of a viscous compressible fluid with variable dynamic viscosity $\eta$ and assuming bulk viscosity coefficient
equal to zero i.e., $\eta_B = 0$. %The bulk viscosity is connected with finite relaxation process rates e.g., nuclear or chemical reactions.
%For a review of fluid mechanics see Landau \& Lifshitz 1959 \cite{DM59:FluidMechaSeconEditiVolumCoursTheorPhysi}.

We consider a fixed volume $V$  wrapped  within a surface $S$. The rate of the fluid mass change contained 
in the volume $V$ is

\begin{equation}
\frac{\partial}{\partial t} \int_V \rho dV,
\label{Vol1}
\end{equation}
where $\rho$ is the density of the fluid. 

In the absence of source/sink for matter, Eq. (\ref{Vol1}) must be equal to the total mass inflow integrated over the surface. 

Assuming that $\textbf{v}$ is the velocity vector of the fluid, the outward mass flow across an element $d\textbf{S}$ is 
$\rho \textbf{v} \cdot d\textbf{S} $. Therefore, the gained mass by the volume $V$ is calculated by integrating over 
the surface $S$. Using the divergence theorem to express surface integral into volume integral, we have,

\begin{equation}
- \int_S  \rho \textbf{v} \cdot d\textbf{S} = - \int_V \nabla \cdot (\rho \textbf{v}) dV,
\label{Volumen2}
\end{equation}
the minus sign is added because we are considering an outflow motion. 

Therefore, equating the above expression to the Eq. (\ref{Vol1}), we find,

\begin{equation}
\frac{\partial}{\partial t} \int_V \rho dV = - \int_V \nabla \cdot (\rho \textbf{v}) dV.
\label{ConservationIntegral}
\end{equation}

From this, the mass conservation law can be written as:

\begin{equation}
\frac{\partial}{ \partial t} \rho + \nabla \cdot (\rho \textbf{v}) = 0.
\label{Conserv1}
\end{equation}

This is the usual Eulerian form of the mass continuity equation.

In order to formulate the conservation law for momentum, it is necessary to define the sources influencing
the variation of momentum. These sources are forces acting on the fluid, they consist in external volume forces  \textbf{${f_{e}}$} and 
internal ones \textbf{$f_i$}. The latter depends on the nature of the considered fluid, i.e., on the assumptions made about the
deformations within the fluid and their relation to internal stresses. For the later, a  \textit{viscous stress tensor} must be introduced,

\begin{equation}
\tau_{ij} = \mu \sigma_{ij} = \mu \left(  (\partial_i v_j + \partial_j v_i)  - \frac{2}{3} (\nabla \cdot v) \delta_{ij}    \right),
\label{ViscousstressTensor}
\end{equation}
where $\sigma_{ij}$ is the \textit{shear tensor} and $\mu$ the dynamic viscosity of the fluid. 

Now, the total internal stresses can be described using the \textit{stress tensor} $\Upsilon$. Denoting by $P$ the pressure, we have,

\begin{equation}
\Upsilon = - P I + \tau,
\label{StressTensor}
\end{equation}
where $I$ is the identity matrix, and $\tau$ the \textit{viscous stress tensor} given by (\ref{ViscousstressTensor}).

Using Equation (\ref{StressTensor}) we can find the internal forces acting on the fluid, 
i.e., $\textbf{f}_i = \nabla \cdot \Upsilon = - \nabla P + \nabla \cdot \tau$.

If the external forces come from a gravitational potential $\Psi$, then we have $\textbf{f}_e = - \nabla \Psi$.

We can now, write down the equation for the momentum conservation, which reads,

\begin{eqnarray}
\rho \left(  \frac{\partial}{\partial t} \textbf{v} + (\textbf{v} \cdot \nabla) \textbf{v} \right)  & = & \nabla \Upsilon + \rho f_e \nonumber\\
									                      & = & - \nabla P  + \nabla \cdot \tau - \rho \nabla \Psi.
\label{ConservationMomentum}
\end{eqnarray}

This equation is commonly referred to as the \textit{Navier-Stokes equation of motion}. For an ideal fluid without internal stresses, 
that is an inviscid flow, i.e., $\Upsilon = 0$,
this reduces to the \textit{Euler equation of motion}.

These set of equations are a general description of fluid motions. Usually they are very difficult to solve and cannot be done without several assumptions about the nature of
the fluid and the geometry of the problem. 

We use cylindrical coordinates ($r, \varphi, z$) to describe our disks. From an analytical (and numerical) point of view this is a very 
natural choice, not just due to the disk geometry but also because integrations are performed from the last stable orbit $r_{lso}$, and not from the origin of the coordinate at $r = 0$. In
that point the Jacobian diverges causing several issues in numerical calculations.

We also take symmetry considerations and vertical integration of variables, to reduce the dimension of the problem and the number of independent variables.
For instance, it is assumed an axisymmetric disk, which means that all quantities are independent of the  $\varphi$ coordinate, then $\frac{\partial}{\partial \varphi} = 0$.
When dropping the azimuthal coordinate, it is no longer possible to resolve local disturbances in the azimuthal direction. This kind of term could be a source
of viscosity and/or turbulence interesting to observe, but  when we deal with a very extended disk ($\sim 150 ~\rm pc$) computational cost in adding the azimuthal component would be prohibitive.  

As for the vertical integration, the idea is to get rid of all $z$ dependencies by integrating the equations through the depth of the disk, so we assumed that
the flow is symmetric with respect to the equatorial plane (mirror symmetry about this plane). This procedure allows us to decouple the vertical and radial directions. In that case
we have for the velocity components $v_r$, $v_\varphi$, and $v_z$,

\begin{equation}
v_r(r,z) \simeq v_r(r), ~ ~ ~  v_\varphi(r,z) \simeq v_\varphi(r),   ~ ~ ~  v_z \simeq 0.
\end{equation}

The condition $v_z = 0$ is, of course, equivalent to say
that the disk is in hydrostatic equilibrium along the z-axis.

Vertical integration is then understood as integration over vertical coordinate with the scale of height of the disk as limits, i.e., $\pm H(r)/2$ . This 
implies that rather than dealing with quantities per unit volume, we will deal instead with quantities per unit 
surface. Integrating the density $\rho$ along the $z$-axis we obtain the surface density given by,  

\begin{equation}
\Sigma(t, r) \equiv \int_{-\infty}^{+\infty} \rho dz ~~ = \int_{-H(r)/2}^{+H(r)/2} \rho dz ~~ = H(r) \rho(t, r, z=0),
\end{equation}
where $H(r)$ is the total scale of height. Like the density $\rho$, all other quantities are vertically averaged. For instance,
for the viscous stress we obtain,

\begin{equation}
T_{\mu \nu} \equiv \int_{-H(r)/2}^{+H(r)/2} \tau_{\mu \nu} dz.
\end{equation}

Using the above symmetry considerations, we can easily derive 
the mass conservation equation, the radial and azimuthal components of the Navier-Stokes equation (\ref{ConservationMomentum}) within cylindrical 
coordinates. Therefore, we have:

\subsubsection{mass conservation}

From Eq. (\ref{Conserv1}), we simply obtain,

\begin{equation}
\frac{\partial \Sigma}{\partial t} + \frac{1}{r} \frac{\partial(r\Sigma v_r)}{\partial r} = 0.
\label{contonuity}
\end{equation}

\subsubsection{the radial component}

The equation for the radial transport, including gas pressure, and the three 
components of viscous stress tensor $T_{\mu \nu}$ (Eqs. (\ref{tau0}), 
(\ref{tau2}), (\ref{tau1})), is given by,

\begin{equation}
\frac{\partial v_r}{\partial t} + v_r \frac{\partial v_r}{\partial r} = -\frac{1}{\Sigma(t, r)} \frac{\partial P}{\partial r} + \frac{v_\varphi^2}{r} -
 \frac{\partial \Psi}{\partial r} + a_r^{visc}.
\label{r-comp}
\end{equation}

Approximations that can be used  without strikingly changing the physics of the system is to neglect gas gradient pressure
along the radial direction, and assume the components $T_{r r} = T_{r \varphi} = 0$, hence $a_r^{visc} = 0$. These assumptions 
are well established and justified due the fact that the azimuthal velocity is greatly superior to the radial 
velocity (i.e., $v_\varphi \gg v_r$). In that 
case, the only important viscous force lies between two consecutive annulus $r - \Delta r$ and $r$ which creates a force per unit
surface $T_{r \varphi}$ and thus exerts a torque which is responsible for transporting the angular momentum. Therefore
the last Equation (\ref{r-comp}) simplifies to,

\begin{equation}
\frac{\partial v_r}{\partial t} + v_r \frac{\partial v_r}{\partial r} =  \frac{v_\varphi^2}{r} -
 \frac{\partial \Psi_T}{\partial r} ,
\label{r-componente}
\end{equation}
where viscous forces play no role in the radial component. The gravitational potential $\Psi_T$ includes the contribution from both,
the central black hole and the disk itself. Note that there is a continued flow of matter as time passes, therefore the 
gravitational potential is also a function of time. We then have  $\Psi_T = \Psi_T(r, t) = \Psi_{bh}(r, t) + \Psi_{disk}(r, t)$.

\subsubsection{the azimuthal component}

The equation for the azimuthal motion includes the viscous forces responsible for the angular momentum transport. From the equation (\ref{ConservationMomentum}) we have,

\begin{equation}
\frac{\partial v_\varphi}{\partial t} + v_r \frac{\partial v_\varphi}{\partial r} + \frac{v_r v_\varphi}{r} = a_\varphi^{visc},
\label{phi-comp}
\end{equation}
where $a_r^{visc}$ and $a_\varphi^{visc}$ are the radial and azimuthal accelerations due to the disk viscosity ($r$ and $\varphi$-component 
of $\nabla \cdot \tau$ in cylindrical coordinates), they are given by 

\begin{eqnarray}
a_r^{visc} & = & \frac{1}{r \Sigma(t, r)} \left(  \frac{\partial (r T_{r r})}{\partial r} - T_{\varphi \varphi} \right) \label{rVisc},\\
a_\varphi^{visc} & = & \frac{1}{r \Sigma(t, r)} \left(  \frac{\partial (r T_{r \varphi})}{\partial r} \right), \label{thetaVisc}
\end{eqnarray}
where the terms $T_{\alpha \beta}$ of the components of the 
vertically integrated viscous stress tensor (\ref{ViscousstressTensor}) are given in cylindrical coordinates, by

\begin{eqnarray}
T_{r r}             & = & 2 \nu \Sigma(t, r) \left[ \frac{\partial v_r}{\partial r} - \frac{1}{3} \nabla \cdot \textbf{v} \right] \label{tau0},\\
T_{\varphi \varphi} & = & 2 \nu \Sigma(t, r) \left[ \frac{v_r}{r} - \frac{1}{3} \nabla \cdot \textbf{v} \right] \label{tau2}, \\
T_{r \varphi}       & = &  \nu \Sigma(t, r) \left[ r \frac{\partial}{\partial r} \left( \frac{v_\varphi}{r}  \right) \right], \label{tau1}
\end{eqnarray}
the divergence of the velocity field in the above equations, can be written as,

\begin{equation}
\nabla \cdot \textbf{v} = \frac{1}{r} \frac{\partial}{\partial r} (r v_r).
\end{equation}

When $\nabla \cdot \textbf{v} = 0$, the fluid is said to be incompressible, which is not our case. We let the fluid be compressed near
the location  where
forces are applied, that is, its density increases locally in response to that force. The
compressed fluid expands against neighboring fluid particles causing the neighboring fluid
itself to compress and setting in motion a wave pulse that travels throughout the system.

The radial acceleration due to the viscosity (Eq. \ref{rVisc}) and the pressure gradient term, present in Equation (\ref{r-comp}) can be neglected, their contributions 
are  negligible next to the gravitational forces in the radial direction. Anyway, we do take into account these quantities in 
our calculations.

\subsubsection{the scale of height
}
%vertical balance
The vertical component of Equation (\ref{ConservationMomentum}) can be greatly simplified if we assume very small velocities  
(static situation $\frac{\partial }{\partial t} = 0$) and no viscous forces in the $z$-direction
(i.e., $v_z \simeq 0$ and $\tau_{zz} = 0$), the disk is being confined to the equatorial plane. Then we can neglect all the left-hand side term of the equation. The remaining terms
to balance are the pressure force in the vertical direction with the gravitational forces (from both the central object and the disk itself). This balance
is called the \textit{hydrostatic equilibrium}, and reads:

\begin{equation}
\frac{1}{\rho} \frac{\partial P}{\partial z} = - \frac{\partial \Psi}{\partial z},
\label{equil}
\end{equation}
where $P$ should be the total (vertically-averaged) pressure composed by

\begin{equation}
P = P_{gas} + P_{tur} + P_{rad},
\label{TotalPresion}
\end{equation}
where  $P_{gas} = \rho c_s^2$, the turbulent pressure $P_{tur} = \rho <v_t^2>$, 
the radiation pressure $P_{adr} = a T^4/3$, and $a = 7.564 \times 10^{-15}$ $\rm erg$ $\rm  cm^{-3}~ K^{-4}$ is the radiation constant.

The above Eq. (\ref{TotalPresion}) can be rewritten as

\begin{eqnarray}
P & = & \rho c_s^2 + \rho <v_t^2> +  \frac{a T^4}{3}, \nonumber\\
 & = & \rho c_s^2 \left\{  1 +  \frac{<v_t^2>}{c_s^2} +   \frac{a T^4}{3 \rho c_s^2}   \right\}, \nonumber \\
&= & \rho c_s^2 \left\{  1 +  \frac{<v_t^2>}{c_s^2}     +\frac{2 H a T^4}{3 \Sigma c_s^2}      \right\}, \nonumber\\
& =& \rho c_s^2 \left\{ 1 + \epsilon^2 + \gamma^2 H \right\}, \label{Hsemi}
\label{TotalPresion1}
\end{eqnarray}
where $\epsilon^2 \equiv {<v_t^2>}/{c_s^2}$, is the ratio between the turbulent velocity and sound speed, 
and $\gamma$ is defined as  $\gamma^2 \equiv {2 a T^4}/{3 \Sigma c_s^2}$.

%% Calculating epsilon %%%%%%%%%%%%%%%%%
To calculate the turbulent velocity $<v_t>$ we assume an 
isotropic turbulence. In that case, the kinematical turbulent viscosity $\nu$ is given by

\begin{equation}
\nu = \frac{1}{3} <v_t> <l_t>,
\label{TurViscos}
\end{equation}
where $<l_t>$ corresponds to the characteristic scale of eddies. The turbulent velocity can be approximated using
$<v_t>  \simeq <l_t> \Omega$. Then, the last equation can be recast as

\begin{equation}
\nu = \frac{1}{3} \frac{<v_t>^2}{\Omega}.
\label{}
\end{equation}

Matching this expression with the used viscosity prescription $\nu $, and isolating $<v_t>$,
we have (using $\Omega = v_\varphi / r)$,

\begin{equation}
<v_t>^2 \simeq  3 \nu \Omega.
\label{}
\end{equation}

Therefore, from the definition of the parameter $\epsilon$  we obtain,

\begin{equation}
\epsilon^2 = \frac{3 \nu \Omega}{c_s^2}.
\label{}
\end{equation}

%%%

Defining the total scale of height by the relation 

\begin{equation}
H = \frac{\rho}{\mid \frac{d\rho}{dz} \mid},
\label{scaleH}
\end{equation}
and using Equation (\ref{equil}) we can obtain the values of $H(r)$ as a function of the radius $r$. To find an adequate expression, notice that we can always write 

\begin{equation}
\frac{dP}{dz} = \frac{dP}{d\rho} \frac{d\rho}{dz},
\label{dpdz}
\end{equation}
therefore, using the definition of the effective scale of height, we obtain

\begin{equation}
H = - \frac{dP}{d\rho} \frac{1}{g_z}.
\label{HcasiFinal}
\end{equation}

The first factor in the rhs of the above equation, can be obtained from Eq. (\ref{Hsemi}), 
giving 

\begin{equation}
\frac{dP}{d\rho} =  c_s^2 ( 1 + \epsilon^2 + \gamma^2 H ),
\label{dP}
\end{equation}
and the second one (i.e. $1/g_z$), were $g_z$ is the vertical component of the gravity acceleration, is taken from  $g_z = - \partial \Psi_T/\partial z = G  M_{bh} z/r^3 + 2 \pi G \Sigma z/H$, in which the total gravitational potential was considered.

With these quantities, Eq. (\ref{HcasiFinal}) transforms into,

\begin{equation}
H = \frac{\rho c_s^2 ( 1 + \epsilon^2 + \gamma^2 H )}{2 \pi G \Sigma  + \Omega_K^2 H}.
\label{HCasiCasiFinal}
\end{equation}

Resolving the above equation for $H$, after some algebra  we find a consistent expression for the scale of height 
involving different pressure sources:

\begin{equation}
H = - \frac{c_s}{\overline{Q} \Omega_K} \left[  (1 - \beta) - \sqrt{(1 - \beta)^2 + \overline{Q}^2 (1 + \epsilon^2)} \right].
\label{Hfinal}
\end{equation}

In this equation, there are two important quantities:  $\overline{Q}$  and $\beta$. 
The first one was defined as:

\begin{equation}
\overline{Q} \equiv \frac{\Omega_K c_s}{\pi G \Sigma},
\label{Qbarra1}
\end{equation}
which gives the limit between a self-gravitating and a keplerian regime for the scale of height. 
Given the similarity, it must not be confounded with the Toomre 
parameter $Q$. The latter considers the epicyclic frequency $k$, while $\overline{Q}$ considers the Keplerian angular velocity. Only in the case 
of Keplerian motion, when $k = \Omega_K$, these values coincide i.e., $Q = \overline{Q}$.

The second quantity was defined as:

\begin{equation}
\beta \equiv \frac{a T^4}{3 \pi G \Sigma^2}.
\label{DefBeta}
\end{equation}

The parameter $\beta$ measures the relative influence of the radiative pressure with respect to the self-gravity of
the disk. When radiation pressure is increased, it can overcome gravity. The disk is then inflated (or it could be destroyed),
producing a slim disk. 

In this model, we do not impose a thin disk condition as in the the case of the Shakura-Sunyaev models 
where $H/r << 1$.
The vertical extent of the disk $H$ is allowed to be $H/r \lesssim 1$. 

Only in the inner region of the disk, where temperatures are high enough and $\beta \gg 1$, 
the scale of height of the disk $H$ could be pumped up by radiation pressure and
be comparable to $r$, i.e., $H \sim r$. This force creates a small hot corona in this region. Moving away from the center, temperatures
drop quickly and gravitational tidal forces overcome largely the radiation pressure.

It is important to mention that radiation pressure must be taken into account only when the disk is optically thick.
In the opposite case, radiation passes through the gas  without exerting any force, hence it is negligible. In
such a case $\beta = 0$. This condition must to be applied in order to calculate the scale of height $H$ from 
Eq. (\ref{Hfinal}).

Two interesting limiting cases can be explored from the equation for $H$ (Eq. \ref{Hfinal}). One corresponds to
the case when the radiation force is negligible ($\beta \ll 1$) and self-gravity dominates the entire disk ($\overline{Q} \ll 1$),
applying these in Eq. (\ref{Hfinal}), we obtain

\begin{equation}
H \simeq \frac{c_s^2 (1 + \epsilon^2)}{2 \pi G \Sigma}.
\label{Limt1}
\end{equation}

When the turbulent velocity $<v_t>$ equals the sound speed, i.e., $\epsilon \simeq 1$, we have a typical expression
for the scale of height for self-gravitating accretion disk, that can be obtained almost directly from
the vertical balance (Eq. \ref{equil}),

\begin{equation}
H \simeq \frac{c_s^2 }{ \pi G \Sigma}.
\label{Limt1-1}
\end{equation}

The other case arises when the radiation force is still unimportant but also, the self-gravity of the disk 
is negligible ($\overline{Q} \gg 1$), in this case the disk is Keplerian and we have,

\begin{equation}
H \simeq \frac{c_s \sqrt{1 + \epsilon^2}}{ \Omega_K}.
\label{Limt2}
\end{equation}

Once the temperature of the disk is calculated from the energy equation, it is possible to immediately evaluate
the expression for $H$; recall that scale of height is a time-dependent function of the temperature  $H = H(t, T)$.  

%%%%%    end of  the scale of height

The system of Navier-Stokes equations still has  to be supplemented with a specification of the kinematic viscosity $\nu$ (related to the dynamical viscosity through $\nu = \mu / \rho$) 
as a function of other flow variables. Using the same symmetry arguments, vertical integration of the kinematic viscosity is defined as,

\begin{equation}
<\nu> = \frac{1}{\Sigma} \int_{-H(r)/2}^{+H(r)/2} \rho ~ \nu ~ dz
\label{kine-averaged}
\end{equation}

For simplicity, when we talk about the kinematic viscosity we mean the \textit{z-averaged} kinematic viscosity given by expression (\ref{kine-averaged}).

The choice of $<\nu>$ is the most obscure and speculative aspect of accretion theory and a key problem to be solved.  The main issue is to find a source of $\tau$ stress that could
be effective to transport angular momentum making accretion possible in a reasonable timescale.  

The next subsection, the $\alpha-$disk model is presented with their main assumptions, features, successes and limitations. %Next, our model is proposed and analyzed

\subsection{Standard model: $\alpha$-disk }

The \textit{standard} model of accretion disk was first formulated in the papers of Shakura  (1972)   and Shakura \& Sunyaev  (1973), and then generalized to the Kerr-metric by Novikov \& Thorne (1973). It is interesting to recall that
all these soviet authors (Shakura, Sunyaev, Novikov), pioneers in the development of the accretion theory, were collaborators of Y.B. Zel'dovich  
whose influence was of extremely importance even if his name does not appear in these classical papers.

The main idea of the $\alpha-$model was to describe a geometrically thin non-self gravitating disk by hydrodynamical equations averaged
over the disk thickness. Geometrically thin and non-self-gravitating disk means that the scale of height of the 
disk $H$ (thickness) is much smaller than the radial distance $r$ (i.e., $H/r \ll 1$) and the mass of the disk $M_d$ 
is much smaller than the mass of the central object $M_\bullet$ (i.e., $M_d \ll  M_\bullet$) so the gravitational influence
of the disk is negligible.

The most important process governing the accretion of rotating matter is the action of viscous stress within the flow. The goal is to describe the 
motion of a viscous compressible fluid with variable dynamic viscosity $\eta$. Viscous stresses drive accretion by transporting angular momentum
outward and mass inward; it is also a way to convert gravitational energy of matter into heat, which will be released toward the top and bottom of the surface
of the disk and radiated away.

The evolution of disks under viscosity was also 
extensively studied in a paper from Lynden-Bell \& Pringle (1974).

The original \textit{standard} disk model (the $\alpha$-disk model) also supposes an optically thick accretion disk and a turbulent
fluid described by a viscous stress tensor which is proportional to the total pressure. It is then parametrized as

\begin{equation}
\tau_{r \varphi} = \alpha \rho c_s^2 = -\alpha P,
\label{parameter1}
\end{equation}
where $\alpha$ is a dimensionless constant, that can be fixed for values between zero (case when accretion is halted)
and close to one,  and $c_s$ the sound speed of the flow. The idea is that viscosity is generated by turbulent motion, and the typical turbulence scale $l_t$ has the same
order of the vertical disk scale $H$. On the other hand, turbulent motion cannot be supersonic, otherwise energy would be dissipated very fast and the typical turbulent
velocity would drop below the sound speed.

Using the definition of the viscous tensor $\tau_{r \varphi}$ (Eq. \ref{tau1}), the last parameterization (Eq. \ref{parameter1}) is completely equivalent to considerer the 
kinematical viscosity $<\nu>$, given by

\begin{equation}
<\nu> = \alpha c_s H.
\end{equation}

This form of the $\alpha-$model immediately reflects the fact that turbulent viscosity is considered proportional to the turbulent velocity $v_t$ and the scales of eddies $l_t$ in the
turbulent pattern, i.e., $\nu \propto v_t l_t$, where these quantities are supposed to be $v_t \sim c_s$ and $l_t \sim H$ as expressed above. 
Since the turbulent velocity must be $v_t \lesssim c_s$ and the eddies scales $l_t \lesssim H$, we obtain two upper limits implying that $\alpha$ should be $\lesssim 1$.

It must be clear that the $\alpha-$model is not a real theory of viscosity in accretion disk it is just a way to hide our ignorance about viscosity, replacing the
unknown parameter $\nu$ by another unknown parameter $\alpha$ which is supposed to be $\lesssim 1$. Nonetheless, the model can be easily compared
with observations constraining even more the values of $\alpha$. For instance, typically quoted values for $\alpha$ range from $\sim 0.01$ for protostellar 
disks (Hartmann et al. 1998) to  $\sim 0.1$ for galactic binaries (Lasota  2001).

The model assumes that the disk is in local thermal equilibrium, does not 
advect heat inwards, and can radiate  the viscous heat efficiently as a blackbody.

The main assumptions in the standard disk model can be summarized as follows:

\begin{itemize}
 \item Gravitation is only determined by the central object (no self-gravity effect)
\item The disk is geometrically thin. i.e., $H / r \ll 1$
\item The disk is steady. i.e., $\partial / \partial t \equiv 0$
\item The disk is axisymmetric i.e., $ \partial / \partial \varphi \equiv 0$
\item Azimuthal motion dominates over radial motion  i.e., $v_\varphi \gg v_r$
\item Hydrostatic balance is assumed in the vertical direction
\item The disk is optically thick in the vertical direction
\item Viscous heat dissipation balance radiation output i.e., $ Q^+ = Q^-$
\item The $r-\varphi$ component of the viscous stress tensor is assumed proportional to the pressure i.e., $\tau_{r \varphi} = - \alpha P$. All other components are zero
\item There are no magnetic fields present in the disk
\end{itemize}

Observations show the electromagnetic energy distribution in frequency band. The success of an accretion disk model lies in the ability to describe, among other phenomena,
these observations. Of course they depend on the nature of the source, whether  the object is a binary system or an active galaxy for instance. 

This model was quickly recognized; it supplies a robust theory of the accretion flow, easy
to apply to observations. 

It successfully fits, for instance, the emission properties of dwarf-novae, X-ray binaries (Cannizzo  1993),   
UV-soft X-ray emission of AGNs (Sun \& Malkan 1986). However, this picture is unstable
under \textit{thermal} and \textit{viscous} instabilities 
(this point will be clarified later), moreover, other viscosity prescription could work as well in order to describe the aforementioned observations.

\subsection{The classical Eddington limit} \label{EddingontLimitClassic}

The mass $M$ of the central object plays an important role in the nature of the accretion disk, especially when considering
the maximum luminosity allowed by the system. This maximum luminosity is called the \textit{Eddington limit}.

The \textit{standard Eddington limit}\footnote{The \textit{standard} or \textit{classical}
Eddington limit, introduced by Sir Arthur Stanley Eddington,
only takes into account electron scattering processes.}, 
was originally  derived for stars. It
imposes a fundamental upper limit for the radiative luminosity of 
a \textit{steady spherical} accretion flow, which is usually called
the Eddington luminosity. The basic idea is to find the equilibrium situation between infalling material when interacts 
with the outward radiation flux. Assuming that the
accreting material is fully ionized hydrogen, radiation pressure acts mainly on the free electrons. The associated force 
over one  electron is due to Thomson scattering and it is
given by $\sigma_T F / c$, where $\sigma_T = 6.7 \times 10^{-25}$ $\rm cm^2$ is the Thomson cross section, $F$ the outward 
flux (in $\rm erg ~s^{-1}~ cm^2$) and $c$ the speed of light.
The radiation pressure pushed up electrons against the gravitational force, at the same time electrons drag the protons 
with them due to Coulomb interaction, but the force
of radiation into protons is negligible next to electrons because the scattering-cross section of protons is 
$(m_e/m_p)^2 \simeq 25 \times 10^{-8}$ smaller than 
for electrons. 

Under this condition, accretion is only possible if the total gravitational attraction overcomes the radiation outward 
force. The Eddington limit corresponds 
to the situation when these quantities are equal. The gravitational force exercised from the central object of mass $M$ 
to a  electron-proton  pair
is given by  $G M (m_e + m_p)/r^2 \simeq G M m_p /r^2$, and must be equal to the radiative force over one electron. Then we have

 \begin{equation}
\frac{G M m_p}{r^2} = \frac{\sigma_T F}{c}.
\label{Edd1}
\end{equation}

The flux $F$ of the accreting source is related to his luminosity through, $F = L / 4 \pi r ^2$ (as mentioned above, the system is spherically symmetric), then
the last equation becomes 

 \begin{equation}
\frac{G M m_p}{r^2} - \frac{\sigma_T }{c} \frac{L}{4 \pi r^2} = 0.
\label{Edd2}
\end{equation}

The luminosity that fulfills this condition is the Eddington luminosity,

 \begin{equation}
L_E =    \frac{4 \pi G M m_p c}{\sigma_T} \simeq 1.26 \times 10^{38} (M/M_\odot)  ~~ \rm erg ~ s^{-1}.
\label{EddLum}
\end{equation}

If we consider that an object (black hole, white dwarf, neutron star, etc.) is only powered by (spherical) accretion, this condition implies a 
limit on the \textit{steady} accretion rate, called the Eddington rate $\dot{M_E}$.  It can be obtained from the Eddington luminosity, 

 \begin{equation}
\frac{G M }{r} \dot{M_E} = L_E   \hspace{0.2cm}  \rightarrow  \hspace{0.2cm}  \dot{M_E} = L_E \frac{r}{GM} = 4 \pi r \frac{m_p c}{\sigma_T}.
\label{EddRate}
\end{equation}

When accretion exceeds this limit, the associated luminosity will exceed $L_E$, then the outward pressure radiation
overtakes gravitational attraction hence; accretion is halted, turning off the source.

For realistic astrophysical models, this approach needs to be reconsidered. Several of the used assumptions could not be true. For example, a mixture of elements other
than hydrogen, or the ionization level of the gas invalidates the latter. Spherical accretion seems more difficult to be produced. Viscous gas tends to form a
disk when falling into a potential well due to loss of angular momentum, therefore, the geometry ceases to be spherical. 

In the case of accretion disks, the Eddington limit should  not be used as exposed in this section. 

%% Make a section about the Eddington Limit !!!
%We shall see a more appropriate approximation for this limit in Chapter %\ref{ModelFinal}, Section \ref{EDLimAccDisk}. 

\subsection{Temperature profile and  disk continuum spectra in $\alpha$-disks}\label{TempProfileInAlphaDisk}

Viscous processes in the internal region of the disk (equatorial plane) converts the potential energy of the accreting matter into thermal energy which is
then completely radiated away throughout the surface of the disk.

Viscous heating per unit volume is given by:

\begin{equation}
 q^+ = \rho \nu \left( r  \frac{ d\Omega_K}{dr}  \right)^2,
\label{qplusvolume}
\end{equation}
where $\Omega_K = (G M /r^3)^{1/2}$ is the keplerian angular velocity.

The radiative flux $F$ in the $z$-direction is given by the diffusion equation -which is only 
valid for optically thick disks-, therefore we have,

\begin{equation}
 F = - D \frac{\partial u}{\partial z},
\end{equation}
where $D = \lambda c /3$ is the diffusion coefficient, $\lambda$ the photon mean free path and $u = a T^4$ correspond to the radiation energy density ($a$ is the 
radiation constant). Notice that from  the radiation energy density, $du = 4 a T^3 dT$, therefore we can rewrite the above equation as (transforming the 
partial derivative into an ordinary one):

\begin{eqnarray}
 F &  =  & - D \frac{d u}{d z} \nonumber \\
   &  =  & - D \frac{d u}{d T} \frac{d T}{d z} \nonumber  \\
   &  =  & - D  4 a T^3 \frac{d T}{d z}\nonumber  \\
   &  =  & - \frac{4 \lambda a c T^3}{3}    \frac{d T}{d z} \nonumber  \\
   &  =  & - \frac{4 a c T^3}{3 \kappa \rho}    \frac{d T}{d z},
\label{RadTranF}
\end{eqnarray}
where we have used for the photon mean free path $\lambda = 1 / \kappa \rho$, with $\kappa$ the Rosseland-mean opacity. 
For high temperatures ($T \geq 10^4$ K) the opacity main sources are electron scattering and free-free 
(or Bremsstrahlung \footnote{German word, 
from \textit{bremsen} ``to brake'' and \textit{strahlung} "radiation", i.e. "braking radiation" or "deceleration radiation")})
absorption (for low metal abundance bound-free should be included),

\begin{equation}
 \kappa = \kappa_{s} + \kappa_{ff} = \kappa_s + \kappa_0 \rho T_c^{7/2},
\end{equation}
where $\kappa_s = 0.4$ $\rm cm^2$ $\rm g^{-1}$ and $\kappa_0 = 0.64 \times 10^{23}$ in cgs units. The opacity of the disk is related to the optical depth through,

\begin{equation}
 \tau = \frac{\kappa \Sigma}{2}.
\end{equation}

Now, from the energy balance equation,   the radiation flux must be compensated by the  viscous dissipation heat, then we have

\begin{equation}
 \frac{\partial F}{\partial z} = q^+.
\end{equation}

Integrating the above equation along the $z$-axis, we simply have

\begin{equation}
 F =  \int_0^{H/2} q^+  dz,
\label{Fqz}
\end{equation}
with the boundary conditions $F(z=0) = 0$, and $F(H/2) = \sigma T_{eff}^4 = q^+ H / 2$ (assuming black body emittance), where $\sigma$ is the 
Stephan-Boltzmann constant, $T_{eff}$ the effective 
temperature, and $H$ the total scale of height of the disk.

From equations (\ref{RadTranF}) and (\ref{Fqz}), we obtain the required equation for the temperature of the disk:

\begin{equation}
 - \frac{4  a c T^3}{3 \kappa \rho}    \frac{d T}{d z}  = q^+ z.
\label{TempEq}
\end{equation}

The above equation means that a local energy balance between dissipated and radiated  energy is assumed, i.e., $Q^+_{diss} = Q^-_{rad}$.

The vertically integrated heat rate per unit surface $Q^+_{diss}$, is obtained from,

\begin{equation}
 Q^+_{diss} = \int_{- \infty}^{+ \infty} q^+ dz = \frac{9}{4} \nu \Sigma \Omega_K^2 = - \frac{3}{2} T_{r \varphi} \Omega_K,
\label{Qplus}
\end{equation}
where the viscous tensor $T_{r \varphi}$ (in a steady situation), is given by,

\begin{equation}
 T_{r \varphi} = - \frac{1}{2 \pi} \sqrt{\frac{GM}{r^3}} \dot{M} \left( 1 - \sqrt{\frac{r_{lso}}{r}} \right).
\end{equation}

Let us recall that in this prescription the term $\nu \Sigma$ is written as:

\begin{equation}
 \nu \Sigma =  \frac{\dot{M}}{3 \pi}   \left( 1 - \sqrt{\frac{r_{lso}}{r}} \right).
\label{nuSigma}
\end{equation}

The radiative cooling is given by;

\begin{equation}
 Q^-_{rad} = 2 \times F = 2 \sigma T_{eff}^4,
\end{equation}
where the factor $2$ represents radiation from the two surface of the disk. 
Notice that the radiative flux of the disk scales as $\propto T_{eff}^4$, a typical characteristic of blackbody emissions.
Using the above equation, along with the condition that $Q^+_{diss} =Q^-_{rad} $, we obtain for the effective temperature:

\begin{equation}
 T_{eff}^4 = \frac{1}{2} Q^+_{diss}.
\end{equation}

Now from equations (\ref{Qplus}) and (\ref{nuSigma}), we obtain the effective temperature $T_{eff}(r)$, which is given by:

\begin{equation}
 T_{eff}^4(r) =  \frac{3 G M \dot{M}}{8 \pi \sigma r^3}   \left( 1 - \sqrt{\frac{r_{lso}}{r}} \right).
\label{EffTemp}
\end{equation}

For  regions far away from the centre (i.e., $r \gg r_{lso}$), the effective temperature scales as:

\begin{equation}
  T_{eff} \propto r^{-3 / 4}.
\label{Tscale3/4}
\end{equation}

The radial dependencies of the effective temperature shows that the surface of the disk is hotter in the inner regions, and the disk is
cooled as one moves away from the centre. Notice also that the temperature at the last stable orbit is equal to,   $T_{eff}(r_{lso}) = 0$.

\subsubsection{Thomson scattering case}

If we assume that the opacity is only  due by Thomson scattering  ($\kappa = 0.4$ $\rm cm^2$ $\rm g^{-1}$), which is a good approximation
for highly ionized disks ($T \gtrsim 2 \times 10^4$ K), 
the temperature profile is obtained by integrating Equation (\ref{TempEq}). Using the relation $a c = 4 \sigma$ and 
the boundary conditions mentioned before, after a simply calculation we obtain:

\begin{equation}
 T = T_{eff} \left[ 1 + \frac{3}{8} \tau \left( 1 - 4 \frac{z^2}{H^2} \right) \right]^{1/4}.
\label{Tprofile}
\end{equation}

The temperature at the equatorial plane is obtained taking $z = 0$ in the above equation, then,

\begin{equation}
 T_c = T_{eff} \left( 1 + \frac{3}{8} \tau \right)^{1/4},
\label{TplaneZ}
\end{equation}
and the optical depth of the disk is given in this case by,

\begin{equation}
\tau = \tau_s = \frac{\sigma_T}{m_H} \rho \frac{H}{2} = \frac{\sigma_T}{2 m_H} \Sigma,
\end{equation}
where $\sigma_T$ is the Thomson cross section, and $m_H$ the proton mass.

Now, replacing the effective temperature $T_{eff}(r)$ (Eq. \ref{EffTemp}) in Equation (\ref{Tprofile}) or (\ref{TplaneZ}) we obtain the
temperature profile or the temperature at the equatorial plane of the disk, respectively.

\subsubsection{Disk spectrum}

We have already discussed that in the classic theory of accretion, the disk is assumed to be 
optically thick, radiating throughout the surface as a blackbody. The blackbody radiation is given 
by the Planck function, 

\begin{equation}
 B_\nu(T(r)) = \frac{2 h}{c^2} \frac{\nu^3}{\exp\{ h\nu/k_B T_{eff}(r) \} - 1}.
\label{PlanckFunction}
\end{equation}

The flux is calculated locally, so at each different radius a different value for the temperature is given.

The spectrum of the disk is then calculated by integrating the specific intensity $B_\nu(T(r))$ over the surface of the disk, hence:

\begin{equation}
 L_\nu = \int_{r_{lso}}^{r_d} B_\nu(T_{eff}(r)) 2 \pi r dr,
\label{SpectroAlpha}
\end{equation}
where $r_{lso}$ is the inner boundary of the disk (last stable orbit radius) and $r_d$ his maximum extension. 

%%%%%%%% replace B(T)

The effective temperature can be assumed to scale as a power-law function:

\begin{equation}
 T_{eff}(r) = T_0 \left( \frac{r_0}{r} \right)^n,
\label{TempPowerLaw}
\end{equation}
where $T_0$ and $r_0$ are some normalization values. 

Replacing the above expression in Eq. (\ref{SpectroAlpha}), and using the Planck function  $B_\nu(T(r))$ (Eq. \ref{PlanckFunction}),
we obtain for the spectral luminosity of the disk:

\begin{eqnarray}
 L_\nu & = & \left( - \frac{r_0^2}{n}\right)      T_0^{2/n} \int_{r_{lso}}^{r_d}  B_\nu(T) T^{-(2/n) - 1} dT, \nonumber \\
      & = &   K_0 \nu^{3 - 2/n}  \int_{r_{lso}}^{r_d} \frac{x^{2/n - 1}}{e^x - 1} dx,
\end{eqnarray}

were $K_0 = \left(  \frac{4 \pi h}{c^2}\right)     \left( \frac{r_0^2}{n} \right)  \left(   \frac{k T_0}{h } \right)^{2/n} $

From the above, we notice that the continuum spectrum scales as:

\begin{equation}
 L_\nu  \propto \nu^{3 - 2/n}.
\end{equation}

Using the theoretical value for the slope $n$ of the temperature in the standard model ($n = 3 / 4$) (Eq. \ref{Tscale3/4}), 
we get for the spectrum $L_\nu \propto \nu^{1 / 3}$. Notice that the calculated spectra is only valid in regions far away from the central object, where
lower temperatures produce low energy photons (i.e., soft X-rays, UV, optical).

These results can be applied to several observations. For instance, X-rays spectra from Nova Muscae 1991 (a black hole binary)  shows soft 
and hard components (Mitsuda et al.   1984). The temperature associated to the soft part can be well fitted within this model, as shown 
below.

\begin{figure}[h]
\centering
\plotone{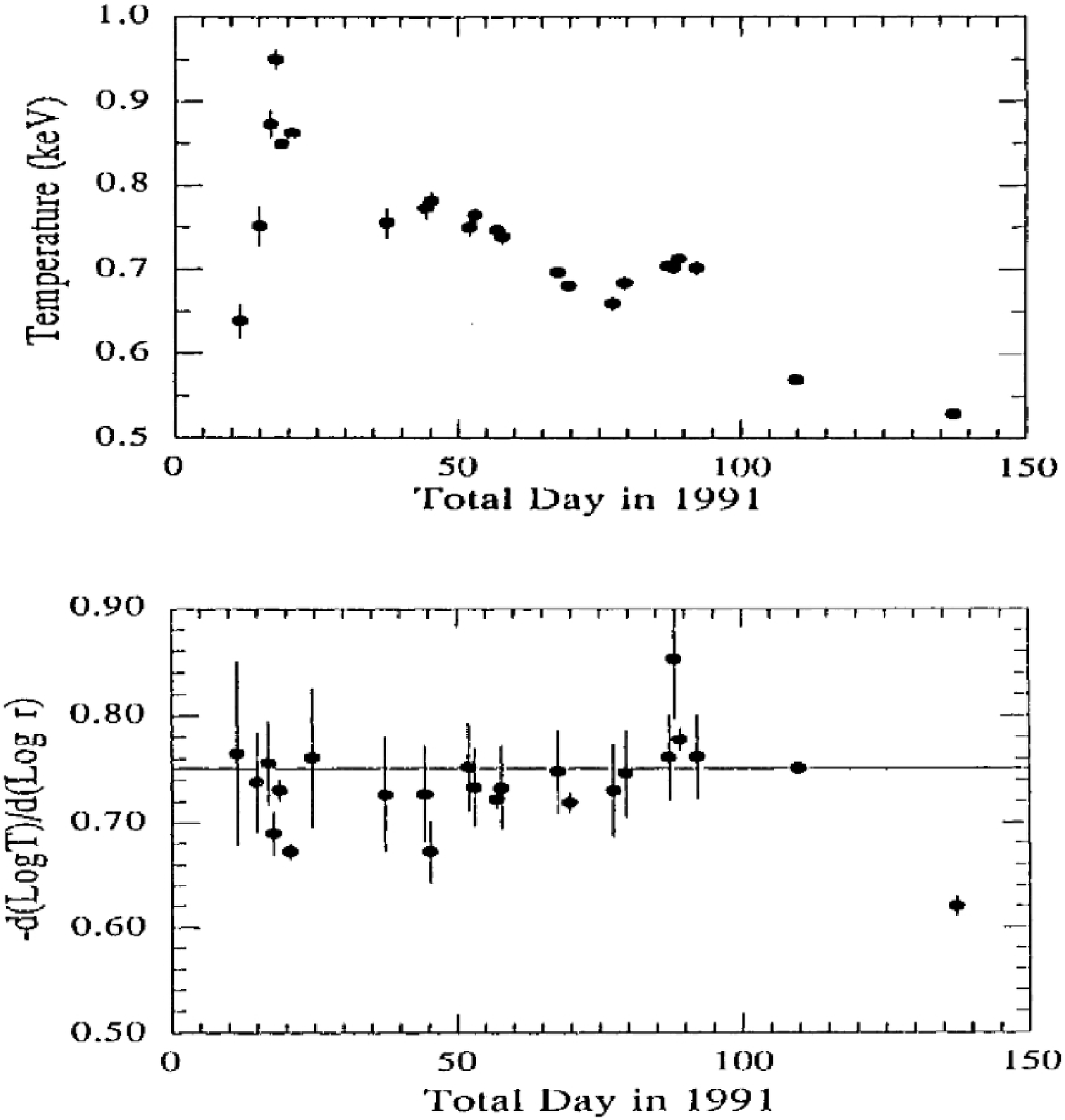}
\caption{Spectral fitting of Nova Muscae 1991 with the standard disk blackbody model. The top 
shows the disk temperature variability. The bottom the temperature exponent. Credits: Mineshige et al. (1994).} 
\label{Mineshige}
\end{figure}

Mineshige et al. (1994)  fitted the soft X-ray spectra of Nova Muscae assuming blackbody emissions with a temperature profile 
$T \propto r^{-n}$. Deriving the temperature gradient from the X-ray data source, they show 
that $-\partial \log T / \partial \log r = n \approx 0.75$ supporting the idea that soft components of the spectra are emitted by a classic accretion disk 
(see Fig. \ref{Mineshige}).

The disk in Nova Muscae is not steady, but its variability is slow enough so that time variations do not affect the emitted
spectra, in fact Cheng et al.   (1992) found that the UV-optical spectrum emissions are  well fitted assuming a steady disk model.

%%%%%%%%%%%%%%%%%%%%%%%%%%%%%%%%%%%%%%%%%% fig 2 %%%%%%%%%%%%%%%%%%%%%%%%%%%%%%%%%%%%%%%%%%%%%%%%%
%%%%%%%%%%%%%%%%%%%%%%%%%%%%%%%%%%%%%%%%%%%%%%%%%%%%%%%%%%%%%%%%%%%%%%%%%%%%%%%%%%%%%%%%%%%%%%%%%%

The bolometric luminosity can be computed from the dissipated energy rate per unit volume $q^+$, therefore:

\begin{equation}
 L = \int_{r_{lso}}^{r_d} q^+ dV = 2 \pi \int_{r_{lso}}^{r_d} q^+ r H dr.
\end{equation}

Using Eq. (\ref{qplusvolume}) and (\ref{nuSigma}), the above equation can be recast as:

\begin{equation}
 L = \int_{r_{lso}}^{r_d} \frac{3 G M \dot{M}}{2 r^2}   \left( 1 - \sqrt{\frac{r_{lso}}{r}} \right) dr = \frac{G M \dot{M}}{2 r_{lso}}.
\end{equation}

The above equation indicates that the disk luminosity is proportional to the accretion rate of the black hole; higher rates, produce more luminous disks. 
Notice also that the factor $GM/2r_{lso}$ correspond to the half of the potential energy difference between infinity to 
the last stable orbit.

Since the accretion rate originates from turbulent viscous processes, the ability to convert potential energy into radiation, comes exclusively from viscous work.

It is worthy to note that, since the disk is hotter in
the inner regions, and the spectra of the disk show essentially blackbody emissions, we conclude that in this model, high energy photons are emitted from the 
inner regions whereas low energy photons come from the outer portions of the disk.

%%%%%%%%%%

\subsection{Uncertainties in $\alpha$-disk models}

As we have seen, Shakura-Sunyaev formalism explains reasonably well disks present in binary
systems but such a model is faced to several stability problems and some assumptions must be revisited.

We begin the analysis by identifying the typical timescales on which the disk structure may vary.

The shortest one corresponds to the dynamical timescale, which corresponds to the orbital timescale

\begin{equation}
t_D = \Omega^{-1},
\label{DynamicalTimescle}
\end{equation}
where $\Omega$ corresponds to the angular velocity, which is deduced from the gravitational potential $\Psi$ from $\Omega^2 = \frac{1}{r}\frac{\partial \Psi}{\partial r}$. For 
a Keplerian, non self-gravitating disk, the angular velocity is given by, $\Omega^2 = \Omega_K^2 = \frac{G M}{r^3}$, where $M$ is the mass of the central object. In this
case, the dynamical timescale is simply given by,

\begin{equation}
t_D = \left( \frac{G M}{r^3} \right)^{-1},
\label{Dynamicam2Timesclae}
\end{equation}
which coincide with the timescale needed for inhomogeneities to be dissipated (be smoothed) in the $z$-direction. In other words, deviations from
hydrostatic equilibrium in the vertical direction are smoothed in a timescale necessary to a sound wave with speed $c_s$, to scroll a distance $H$. This time is given by,

\begin{equation}
t_z = \frac{H}{c_s} = \frac{H}{H \Omega} = \Omega^{-1} = t_D,
\label{VerticalTimescale}
\end{equation}
where we have used the fact that in a thin standard disk, the sound speed  is related to the  scale of height as, $c_s = H \Omega$.

The dynamical timescale can be compared with the viscous timescale in order to estimate the effect of viscosity on the evolution of accretion disk. The viscous timescale gives
the timescales in which matter diffuses through the disk over a distance $r$, under the effect of viscous torques. This quantity is given by,

\begin{equation}
t_v \approx \frac{r^2}{\nu},
\label{ViscousTimeScle}
\end{equation}
inserting the $\alpha$-ansatz in this equation, we obtain

\begin{equation}
t_v \approx \frac{r^2}{\alpha c_s H}.
\label{ViscousTimeScle2}
\end{equation}

Another important quantity is the thermal timescale, which gives the characteristic time needed for the system to recover thermal equilibrium (when $Q^+ = Q^-$).
The thermal timescale is given by a ratio between the thermal energy content ($\sim \rho k T / \mu m_H$ $\sim$ $\rho c_s^2$) per unit surface area and the viscous dissipation
rate ($Q^+ \approx r H \tau_{r \varphi} d\Omega/dr  \approx (9/4) \Omega^2 \nu \Sigma$) per unit surface area, then,

\begin{equation}
t_{th} \approx \frac{c_s^2}{v_\varphi^2} \frac{r^2}{\nu} = M^{-2} t_v,
\label{ThermaTimeScle}
\end{equation} 
where $M = v_\varphi/c_s$ is the Mach number  and $t_v$ the viscous
timescale given by Equation (\ref{ViscousTimeScle}). From the condition imposed in this model, that the viscous energy dissipation is locally balanced by radiation from 
the two disk surfaces, the cooling timescale, which is given by the ratio of the thermal energy content by the radiative lost rate ($Q^- = \sigma T_c^4 / \kappa_R \rho H^2$, when the disk is
optically thick) should be the same as the thermal timescale. When advection dominates the disk, this equality does not hold, and the cooling time will be
greater than the thermal timescale. 

Now, we can obtain some relations between these characteristic timescales.
We have already  found that $t_D = t_z$, and $t_{th} \approx M^{-2} t_v$. From the viscous timescale (Eq. \ref{ViscousTimeScle2}), we can rewrite
$t_v \approx \frac{r^2}{\alpha c_s H} = \frac{1}{\alpha} \frac{r}{H} \frac{r}{v_\varphi} \frac{v_\varphi}{c_s}$, then,

\begin{equation}
t_v \approx \alpha^{-1} M^2 t_D.
\label{Relation2}
\end{equation}

Summarizing the results, we see that,

\begin{equation}
\left( t_D = t_z \approx \alpha t_{th} \approx \alpha \left( \frac{H}{r} \right)^2 t_v \right)_\Omega,
\label{Relation3}
\end{equation}
and considering that $\alpha \lesssim 1$, the hierarchy for numerical values is

\begin{equation}
t_D = t_z \lesssim t_{th} \ll t_v.
\label{Hierarchy}
\end{equation}

It is common to see astrophysical models based on steady accretion $\alpha$-disks. Generally, these models teach us important features of these objects.
A steady solution means that the system is, in fact, in an equilibrium
situation, and must stay unchanged under small perturbations; otherwise, the steady solution cannot exist in reality.

We will see briefly two cases of instability situation in models based on Shakura-Sunyaev formalism, namely the \textit{thermal} and \textit{viscous} instability.

The \textit{thermal} instability occurs when the energy balance between radiative cooling and viscous dissipation is perturbed, and this perturbation 
grows on a timescale $t_{th}$ much less than the viscous timescale $t_v$, they take place specially in the inner region of (standard) accretion disk, where temperatures
are higher. More clearly, suppose that the disk is in local energy balance ($Q^+ = Q^-$) and we increase the central
temperature $T_c$ by a small perturbation $\delta T_c$, what happens in this model is that, after the perturbation, the local heating and cooling rates increase (which is normal),
but the local heating rate increases much faster than the cooling rate. This will lead to a thermal runaway, making the steady situation unrealistic.

To easily see how these heating and cooling rates change differently, we consider a simple example: Suppose an optically thick and geometrically thin accretion 
disk (i.e., the $\alpha-$ model), where the cooling rate is given by $Q^- \propto T_c^4$ and the dissipation rate is given by 
$Q^+ \sim r H \tau_{r \varphi} d\Omega/dr$. Considering the parameterization
$\tau_{r \varphi} = - \alpha P$ (Eq. \ref{parameter1}) and a radiation pressure $P \propto T^4$, in a vertical hydrostatic equilibrium with $H \sim 2 P / \Omega_K \Sigma$, we obtain
for the heating rate $Q^+ \propto T^8$, while for the cooling rate $Q^- \propto T^4$. Comparing these two expressions, we notice that when $T$  increases, the 
cooling rate will grow more slowly than $Q^+$. Hence, any perturbation in the temperature will lead to a growing instability.

Unless very specific and often unrealistic conditions are supposed, instabilities cannot be avoided.

The general condition for thermal stability can be written as,

\begin{equation}
\left( \frac{d \ln Q^-}{d \ln T} \right)_{\Sigma, \Omega} < ~ \left( \frac{d \ln Q^+}{d \ln T} \right)_{\Sigma, \Omega},
\label{ThermalStability}
\end{equation}
whenever the above inequality is satisfied the system shows a \textit{thermal instability}.

\begin{figure}[h]
%\centering
\begin{center}
\plotone{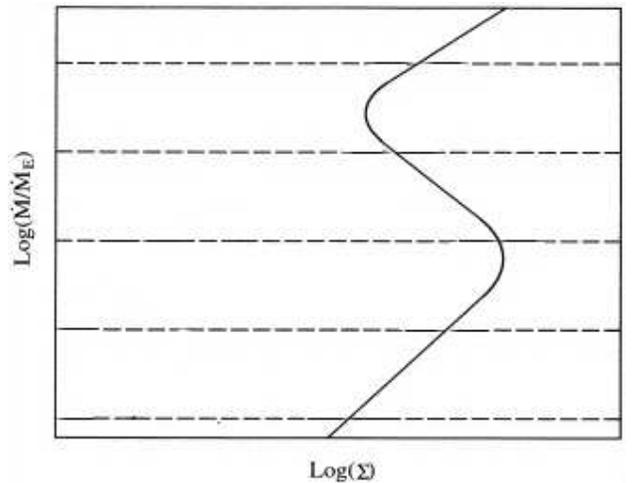} %\vfill
\end{center}
\caption{Schematic S-shape plot with the instability occurring on the intermediate branch of negative slope.}
\label{S-shape}
\end{figure}

If we consider changes in the disk structure, it can be shown that a perturbation $\delta \Sigma$ in the steady solution $\Sigma_0 $ (i.e., $\Sigma = \Sigma_0 + \delta \Sigma$),
the the viscosity perturbation $\delta \mu$ (obtained from $\mu = \nu \Sigma$), obeys a diffusion equation, and 
whenever the condition $\partial \mu / \partial \Sigma > 0$ is locally satisfied,  mass will be removed from that region, leaving a fragmentation of the
disk breaking up in a series of rings. A general condition for disk fragmentation is given by,

\begin{equation}
\frac{\partial \ln \dot{M}}{\partial \ln \Sigma} < 0,
\label{ViscousStability}
\end{equation}
where $\dot{M}$ is the accretion rate. In standard accretion models, the relation $\dot{M} = \dot{M}(\Sigma)$ takes the S-shape, which indicates an instability region 
whenever the slope of the curve is negative (see Figure \ref{S-shape}).

%%%%%%%%%%%%%%%%%%%%%%%%%%%%%%%%%%%%%%%%%% fig 1 %%%%%%%%%%%%%%%%%%%%%%%%%%%%%%%%%%%%%%%%%%%%%%%%%
%[hbtp]
%%%%%%%%%%%%%%%%%%%%%%%%%%%%%%%%%%%%%%%%%%%%%%%%%%%%%%%%%%%%%%%%%%%%%%%%%%%%%%%%%%%%%%%%%%%%%%%%%%

A similar shape emerges in the analysis of $T = T(\Sigma)$, showing the development of thermal instabilities starting from relatively 
low temperatures (e.g. $T > 10^4$ K).

For a complete stability analysis on Shakura-Sunyaev based models, see Piran (1978), and Pringle (1981).

Conclusions arising from stability analysis, is that steady-accretion disk with the $\alpha-$prescription cannot be really stationary. Another
uncertainty in these models is about the type of pressure that should be used in the anomalous stress tensor (Eq. \ref{parameter1});
It should be used only the gas pressure? radiation pressure? something else?
Even just considering gas pressure, and hence avoiding thermal and viscous instabilities, radiation pressure is still important. In fact, in Shakura-Sunyaev models
around black holes, radiation pressure is always dominant for high accretion rates (when the luminosity is near the Eddington luminosity), so it should be included,
but their inclusion leads to thermal instabilities.

%As a conclusion we see the necessity of a non-stationary accretion disk

\subsection{Beyond the standard model} \label{BeyonAlpha}

We have already seen some drawbacks in the $\alpha$-parameterization. It is clear now that a key ingredient to describe the accretion process is
the turbulent  viscosity. It causes the angular momentum loss and the inward motion of the material 
(as well as an outward flow by the external region of the disk). 

Shakura-Sunyaev parameterization remains an ad-hoc description limited to thin disk geometry and negligible disk masses. Now, if you want to 
``construct'' a supermassive black hole through an accretion disk in timescales less than 1 Gyr, needed for consistence with observations of most distant quasars at redshifts $z \sim 6$,  you will need to consider a self-gravitating initial disk.
We expect from this a high accretion rate, that could lead to a super-Eddington regime, and probably an inner region with disk thickness $H(r)$ comparable to the disk size i.e., $H \sim r$. 

For the reasons mentioned above, we need a more 
realistic description of the turbulent nature of viscosity, and relax the assumption that the disk can only be geometrically thin. 

Understanding the origin of turbulent transport is one of the major problem in accretion physics. The nature of the physical
mechanism which causes turbulent motion is still an open question. A good candidate that partially explain turbulent stress is 
the magneto-hydrodynamic instability (MHD instability). It was 
originally discovered by Velikhov  (1959). From this, several authors explored this possibility, and applied they 
results to accretion disk theory.

In these models, the source of turbulent viscosity may arise from magnetic fields, which are generated and
maintained in the disk by a dynamo mechanism. This mechanism could produce local instabilities generating turbulence and hence 
an $\alpha-$effect (with a \textit{magnetic pressure}), similar to those produced by the standard model 
(see for instance Brandenburg  (1995), Balbus \& Hawley  (1998). 
Numerical simulations  showed that a weak magnetic field in accretion disks could lead to a 
very effective transport of angular momentum and hence, an efficient accretion rate for a sufficiently ionized disk flow (e.g.,
Hawley  (1995)).

These models work quite well, they describe very easily 
low-mass X-ray binaries, or systems when 
the central object is actually a magnetic neutron star or a non-magnetic simple star. It is important to note that in those
cases the disk is probably heated by irradiation due the central object, ensuring a sufficiently ionization rate.

The situation is rather different when a black hole is considered as the central source, 
and even more radical if we attempt to describe large scale disks ($\sim$ hundreds parsec radius), where it is expected that the outer region of the disk the
temperature is sufficiently low to form a large amount of molecular gas. Then,
we cannot always expect a high ionization rate in the gas and therefore the necessary conditions to anchor a magnetic field in it.  
 
On the other hand, due to the turbulent nature of the flow, it is expected
that the number of Reynolds be relatively high. This number is defined as,

\begin{equation}
\Re = \frac{\rho u l}{\mu} = \frac{V l}{\nu},
\label{ReNumber}
\end{equation}
where $u$ is the characteristic mean fluid velocity, $l$ the characteristic dimension, and $\nu$ the kinematical viscosity (related to the dynamical
viscosity through $\nu = \mu / \rho$ measured in $\rm m^2$ $\rm s^{-1}$).

Lyndell-Bell and Pringle (1974) pointed out that if an accretion disk is driven by molecular viscosity (where $\nu \rightarrow \rm small$),
the Reynolds number becomes high (see Eq. \ref{ReNumber}), then turbulent eddies develop in the flow, enhancing in turn
the effective viscosity (turbulent viscosity), as a consequence the Reynolds number will decrease. This process will continue until a critical Reynolds number $\Re_c$
is reached, which indicates the onset of turbulence. In this sense, the flow is expected to be self-regulated and characterized by
a critical Reynolds number. Below $\Re_c$ the flow is laminar, but for higher values the flow becomes turbulent.
This ensures that if the angular
momentum transfer is inefficient, turbulence decays triggering instabilities which regenerates 
the turbulence.

This approach was revisited by de Freitas Pacheco \& Steiner (1976 ), who modelled a steady
accretion disk able to explain the observed X-ray spectrum of the binary source Cygnus X-1.
Their prescription is robust, assuring high accretion rates and making magnetic fields needles. 

Notice that a similar treatment was discussed by Duschl, Strittmatter \& Biermann  (1998) (see also, Richard \& Zahn  (1999)), who have introduced the ``$\beta$-viscosity'', which has essentially the
same parametrization as that of de Freitas Pacheco \& Steiner (1976).

The best connexion with experimental data comes from laboratory experiments on rotating Couette-Taylor 
flow (Wendt  1933), Taylor  1936), consistent in a fluid between two cylinders rotating at different 
speeds aiming to study hydrodynamical instabilities in the flow.

Considering the Reynolds number definition (Eq. \ref{ReNumber}), we assume that the characteristic mean fluid velocity is very close to the azimuthal velocity $u \simeq v_\varphi$ and
the characteristic dimension of the system is $l \simeq 2 \pi r$. After the system has adjusted itself to a critical Reynolds number $\Re_c$, we get,

\begin{equation}
\nu = \frac{2 \pi}{\Re_c} v_\varphi r = \beta \Omega r^2,
\label{ViscoBeta}
\end{equation}
where $\Omega = v_\varphi / r$ corresponds to the angular velocity, and the free parameter $\beta \equiv 2 \pi / \Re_c$ gives a 
measure of how turbulent the fluid will be. 

Based on hydrodynamical experiments, common values for the critical Reynolds number are $\Re_c \sim 10^2 - 10^4$. 
Recall that the smaller $\Re_c$ is, the earlier the
turbulence is generated, and then more turbulent the 
fluid is, ensuring efficient angular
momentum transport outside the disk, and allowing material to flow in.

This parametrization of viscosity differs in many aspects from Shakura-Sunyaev models. It allows an efficient transport 
of angular momentum like his predecessor, but 
under the analysis of stability given by Piran  (1978), it shows a stable disk for all $\beta$ values and for several cooling mechanism. For instance, see
Montesinos \& de Freitas Pacheco (2011), were using this 
parametrization they studied the evolution of self gravitating
disks around supermassive black holes  solving numerically the complete set of hydrodynamical equations discussed in this paper.

\section{Evidence of self-gravitating accretion disks}

In the Universe there are several classes of self-gravitating disks. They can be small circumplanetary disks 
with less than $1~ AU$ radius, or protostellar disks with intermediate radius of $(100 - 1000 ~ AU)$, 
or several hundred parsec scale disk as found in AGNs. In all of them, self-gravity plays an important role 
in the physical properties of these object.

After the study of 47 objects, 
Hillenbrand et al.  (1992) argued that Herbig Ae/Be stars\footnote{A young star ($< 10^6 $ yr)
of spectral type A or B, with mass between $1.5$ to $10$ $M_\odot$} 
are surrounded by circumstellar accretion
disks with masses up to six times higher than the mass of the central object.

\begin{figure}[h]
%\centering
\begin{center}
\plotone{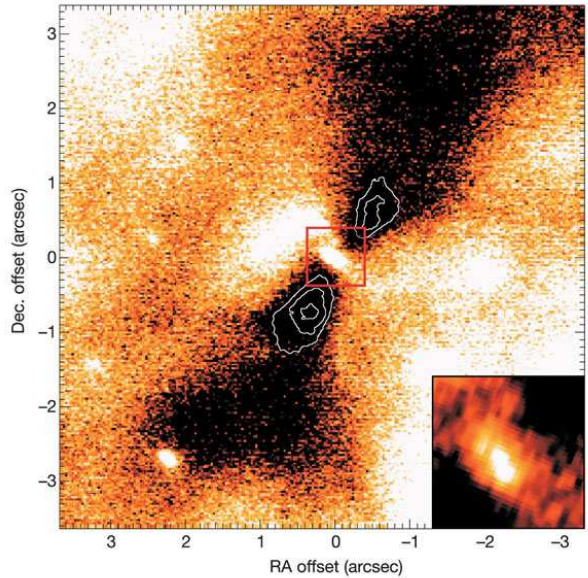} %\vfill
\end{center}
\caption{Silhouette of the accretion disk with a diameter of 20 000 AU, seen against the light from the H II region. White 
contours delineate the densest part of the disk (inner torus). 
Credits: Chini et al. (2004).}
\label{Chini}
\end{figure}

As another example of self-gravitating circumstellar accretion disk, 
Chini et al.  (2004) reports observations of a massive protostar of about $\sim 20 M_\odot$ 
fed by an accretion disk of about $\sim 100 M_\odot$ (see Fig. \ref{Chini}).

%masive disk chini+others2004
%%%%%%%%%%%%%%%%%%%%%%%%%%%%%%%%%%%%%%%%%% fig 2 %%%%%%%%%%%%%%%%%%%%%%%%%%%%%%%%%%%%%%%%%%%%%%%%%
%[hbtp]
%%%%%%%%%%%%%%%%%%%%%%%%%%%%%%%%%%%%%%%%%%%%%%%%%%%%%%%%%%%%%%%%%%%%%%%%%%%%%%%%%%%%%%%%%%%%%%%%%%

We are interested in accretion disks in AGNs. Observational evidence of these objects came from $\rm H_2O$ maser emission
at radius of about $\sim 1~ $pc from the central black hole. From these emissions, the rotational velocity curve of the disk can be
deduced, which differs from Keplerian rotation. 

\begin{figure}[h!M]
%\centering
\begin{center}
\plotone{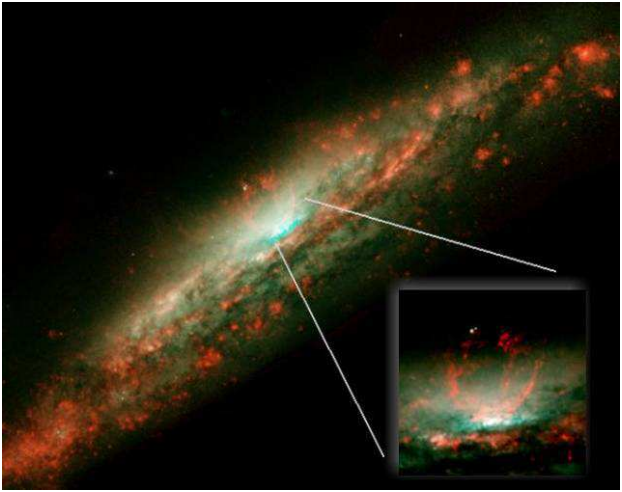} %\vfill
\end{center}
\caption{Galaxy NGC 3079. The structure is more than 3000 light-years wide and rises 3500 light-years above the galaxy's disk.
he smaller photo on the right is a close-up view of the bubble. Astronomers suspect that the bubble is being blown by \textit{winds} 
released during a burst of star formation.
Source: Hubble Space Telescope.}
\label{NGC3079}
\end{figure}

A good example is the case of NGC 3079, classified as a Seyfert 2 (Ford et al.  (1986), Ho et al.  (1997)
(Figure \ref{NGC3079} shows a picture of the whole galaxy). From water maser emissions,  
the obtained disk rotation curve is relatively flat (Sofue 1997). 
They also indicates a central compact object with a 
mass of about $\sim 2 \times 10^6 M_\odot$ enclosed within a radius of $\sim 0.4$  pc.
CO(1-0) observations suggest that the disk surrounding the black hole
is relatively thick; with an aspect ratio in the range of
$H/r \sim 0.1 - 0.5$, and   a  mass of the parsec-scale disk  of about $\sim 3 \times 10^8 M_\odot$, therefore  self-gravitating  (Koda et al  2002). 

Based on stability, cooling, and timescale considerations, Kondratko et al.  (2005) also argue that 
the disk is self-gravitating and clumpy with the necessary conditions
for star formation.

%%%%%%%%%%%%%%%%%%%%%%%%%%%%%%%%%%%%%%%%%% fig 3 %%%%%%%%%%%%%%%%%%%%%%%%%%%%%%%%%%%%%%%%%%%%%%%%%
%[hbtp]
%%%%%%%%%%%%%%%%%%%%%%%%%%%%%%%%%%%%%%%%%%%%%%%%%%%%%%%%%%%%%%%%%%%%%%%%%%%%%%%%%%%%%%%%%%%%%%%%%%

\begin{figure}[h!H]
%\centering
\begin{center}
\plotone{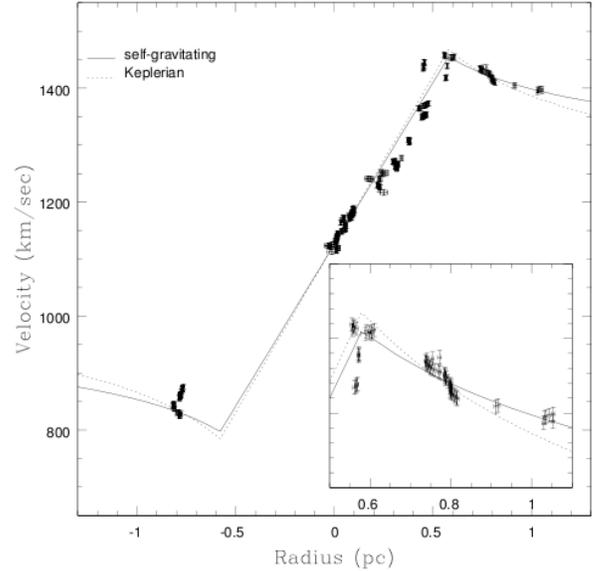} %\vfill
\end{center}
\caption{Fit to the rotation curve from  water maser emission by a self-gravitating accretion disk model.
The small panel shows a blow up of the declining part of the rotation curve together
with the best fit obtained by assuming Keplerian rotation. Credits: Lodato \& Bertin (2003).} 
\label{NGC1068Velocity}
\end{figure}

A similar situation is observed in NGC 1068, another active galaxy classified as a Seyfert 2 that harbor a SMBH. 
From water maser emission observations,
Greenhill \& Gwinn   (1997) show a
non Keplerian rotation disk curve. The velocity field indicates that the rotational velocity
appears to rise steeply in the center toward the peak on the molecular ring  (Fig. \ref{NGC1068Velocity}),   Lodato \& Bertin  (2003)
provide an interpretation of the non-Keplerian rotation in NGC 1068 by means of a self-gravitating accretion disk model 
where the resulting fit-curve is not a power-law (Bertin \& Lodato 1999). They 
also derive a black hole mass of $(8.0 \pm 0.3)\times  10^6 M_\odot$.

More recent observations from the VLTI suggest that the central SMBH is  
surrounded by a torus of gas and dust of size $\sim 3.4 ~\rm pc$ (Jaffe et al.  2004).
It is expected that this torus-like structure should be self-gravitating.

%%%%%%%%%%%%%%%%%%%%%%%%%%%%%%%%%%%%%%%%%% fig 4 %%%%%%%%%%%%%%%%%%%%%%%%%%%%%%%%%%%%%%%%%%%%%%%%%
%[hbtp]
%%%%%%%%%%%%%%%%%%%%%%%%%%%%%%%%%%%%%%%%%%%%%%%%%%%%%%%%%%%%%%%%%%%%%%%%%%%%%%%%%%%%%%%%%%%%%%%%%%

Self-gravity not only affects the rotational curve of the disk, it also influences its emitted spectrum.
For instance, T-Tauri stars\footnote{The last stage of stellar evolution before the main sequence. they are
characterized by emission lines, rapid variability, and X-ray emission, with masses of roughly 0.2 to 5 $M_\odot$}
are objects believed to be surrounded by a disk of gas and dust. The evidence for such disks
comes from the excess of strong infrared  (Rydgren, Strom \& Strom 1976) 
and millimetric radiations (Beckwith et al.  1990), indicating that these emissions
come from  regions far from the central star. 

The light curve in T-Tauri stars can be described by a power law model in the form $L_\nu \propto \nu^{3/2}$  (Adams et al. 1988).
If we take a disk, whose temperature follows a power law $T \propto r^{-\alpha}$, the associated luminosity
takes the form $L_\nu \propto \nu^{4 - 2/\alpha}$ (see Section \ref{TempProfileInAlphaDisk}), therefore
the temperature must decrease with the radius in the form $r^{-1/2}$ to be in agreement with observations of the spectrum. 
The problem arises when assuming a standard $\alpha$-disk, which implies a Keplerian rotation curve.
The temperature profile is obtained when the viscous dissipation equals 
the radiative dissipation i.e., $Q^+_{diss} = Q^-_{rad}$ (assuming black body emissions), making
the temperature decrease as $r^{-3/4}$  contrary to observations. 
In order to explain T-Tauri spectrum, we need another
"non-standard" disk model. A deviation from standard models that fit observational data quite well, is to assume that the 
rotational curve of the disk is not Keplerian. For instance, a T-Tauri star with flat spectrum, could be explained
with a constant rotational curve. This kind of shape appears naturally in Mestel's self-gravitating disk models 
(Mestel  1963). Following this reasoning, the infrared excess could be a consequence of the self-gravity
in the disk. Assuming a marginally stable disk, with Toomre parameter $Q \simeq 1$, gravitational instabilities provide
a supplementary contribution to the heating of the disk, which coincides with the infrared excess (see for instance 
Bertin \& Lodato (1999), Lodato \& Bertin  (2001)).

\begin{figure}[h!H]
%\centering
\begin{center}
\plotone{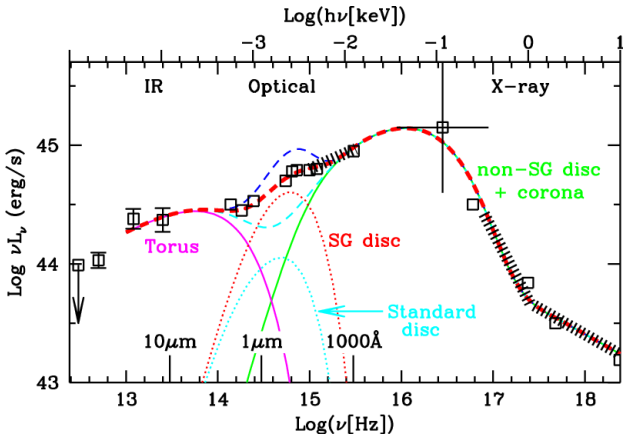} %\vfill
\end{center}
\caption{Mid-IR to X-ray spectral modeling of Ton S 180. The open (white) squares and thick striped lines are the observed data.
The two solid lines (green and magenta) are the spectral components of the dusty torus (left) and the non self-gravitating disk.
The total spectra of those three regions are indicated by dashed lines: the thick line (red) is the spectrum with the self-gravitating
disk, while the case with a standard disc is indicated by a lower curve. Credits: Kawaguchi et al. (2004).} 
\label{TONS180}
\end{figure}

The spectral properties of AGNs are modified in the same way when self-gravity is taken into account. 
Kawaguchi et al.  (2004), explain the origin of Mid-Infrared to X-ray spectrum
emissions of the Narrow-Line Seyfert I galaxy TONS 180, assuming that the disk can be divided in three different zones. 
An inner region of about $\sim 3000 r_s$ from
the black hole, where the disk is slim\footnote{Accretion disk model introduced by
Abramowicz et al.  (1988)}. An intermediate self-gravitating region, where gravitational instabilities and 
the non-Keplerian rotational velocities play an important contribution to viscous heat, and an external region composed
of a dusty torus-like structure, probably at the origin of infrared radiation. Their considerations show 
a very good fit with observations (see Figure \ref{TONS180}), establishing clearly the importance of self-gravity
in AGNs models.

%%%%%%%%%%%%%%%%%%%%%%%%%%%%%%%%%%%%%%%%%% fig 5 %%%%%%%%%%%%%%%%%%%%%%%%%%%%%%%%%%%%%%%%%%%%%%%%%
%[hbtp]
%%%%%%%%%%%%%%%%%%%%%%%%%%%%%%%%%%%%%%%%%%%%%%%%%%%%%%%%%%%%%%%%%%%%%%%%%%%%%%%%%%%%%%%%%%%%%%%%%%

\subsection{Self-gravitating disk model} \label{SGDM}

As we have seen, self-gravity plays an important role in the evolution of astrophysical disks, however 
not many simulations in accretion disk include it. 
This could be due to the difficulty to solve properly the Poisson equation or
the belief that self-gravity could not be important if the mass of the disk $M_d$ is of the same
order or just a few percent higher than the mass of the central object $M_{\bullet}$.
This last can be true if the disk is  very light, but actually it
is not the ratio between the mass of the central object and the disk mass that needs to be compared 
in order to include or exclude the self-gravity from a model, but 
the ratio of their gravitational potential.

At the present time, there is no  simple analytical expression for the gravitational field created by a disk. The best tools are
numerical calculations.

We need a mathematical model for the self-gravity of a disk. Therefore, we need to solve Poisson's equation.
They are several ways to solve this equation in order to compute the gravitational field $\Psi(\bf{x})$ created by a mass
distribution of density $\rho(\bf{x})$, where $\bf{x}$ is a space-vector. But this does not mean that the task is easy. It is
specially hard in astrophysical problems, where the matter distribution is not completely known. This is why,
several assumptions, based on observations, must be considered.

Generally speaking, Poisson's equations is,

\begin{equation}
\nabla^2 \Psi(\bf{x}) = 4 \pi G \rho(\bf{x}),
\label{PoissonEq}
\end{equation}
where $\Psi(\bf{x})$ its the gravitational potential created by $\rho(\bf{x})$, and $G$ is the gravitational constant. 

The most general solution for this equation takes the form,

\begin{equation}
\Psi(\bf{x}) = - G \int_V \frac{\rho(\bf{x'}) d^3\bf{x'}}{\mid \bf{x} - \bf{x'} \mid},
\label{PoissonSol1}
\end{equation}
and  the integration is computed within the volume $V$  enclosing the density distribution $\rho(\bf{x'})$.

The acceleration $\bf{g(x)}$ produced by the gravitational potential $\Psi(\bf{x})$, can be computed from the 
relation $\bf{g(x)} = - \nabla \Psi(\bf{x})$. Then we have,

\begin{equation}
\bf{g(x)} = - G \int_V \frac{\rho(\bf{x'}) (\bf{x}  - \bf{x'} )  d^3\bf{x'}}{\mid \bf{x} - \bf{x'} \mid^3}.
\label{Accel1}
\end{equation}

Using symmetry arguments, and galaxy observations (like observed rotational curves)
we can assume, for simplicity, an  axisymmetric model (i.e., $\partial/\partial \varphi = 0$). This implies that the above space-vector $\bf{x}$, refers to ${\bf x} = (r, \varphi, z)$. 

\begin{figure}[h]
%\centering
\begin{center}
\plotone{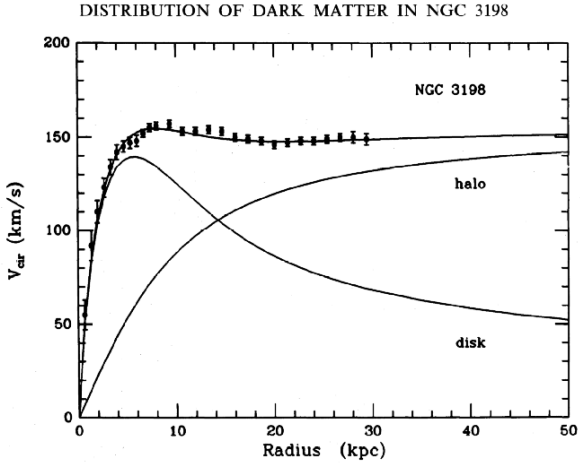} %\vfill
\end{center}
\caption{The rotation curve for the galaxy NGC3198. Note that in the inner region, observations fits very well with the 
disk model.}
\label{RotationNGC 3198}
\end{figure}

Next, we consider the gravitational potential of an
homogeneous spheroid with columnar density profile $\Sigma(r)$ and then make the necessary
limits and extrapolations to obtain the potential on a flat disk approximation. 
These calculations are based on the work of
Wyse and Mayall  (1942); E.M. Burbidge, G.R. Burbidge and Prendergast (1959), 
whose purpose was to study the relation between the density distribution on disks (or geometries close to it) and their 
gravitational field, aiming to reconstruct rotational curves in the nuclear region (inner region where dark matter plays no appreciable role
$r < 3$ Kpc, see Figure \ref{RotationNGC 3198})
from several observed galaxies like, M31, M33, spiral nebula NGC 2146, NGC 3198, amount others. 
Their analysis gives a recipe on how to calculate Newtonian gravitational potentials produced by some density distributions 
in a disk geometry.

%%%%%%%%%%%%%%%%%%%%%%%%%%%%%%%%%%%%%%%%%% fig 5 %%%%%%%%%%%%%%%%%%%%%%%%%%%%%%%%%%%%%%%%%%%%%%%%%
%[hbtp]
%%%%%%%%%%%%%%%%%%%%%%%%%%%%%%%%%%%%%%%%%%%%%%%%%%%%%%%%%%%%%%%%%%%%%%%%%%%%%%%%%%%%%%%%%%%%%%%%%%

\begin{figure}[b]
%\centering
\begin{center}
\plotone{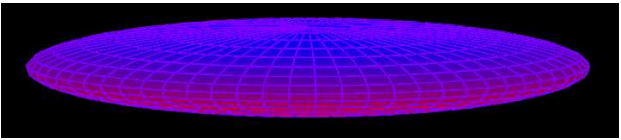} %\vfill
\end{center}
\caption{Oblate spheroid.}
\label{oblate}
\end{figure}

We start our analysis assuming an oblate spheroid (Fig. \ref{oblate}) with major and minor radii $a$ and $b$ respectively, 
eccentricity $e = (1 - b^2/a^2)^{1/2}$ and a density $\rho(r)$.
%%%%%%%%%%%%%%%%%%%%%%%%%%%%%%%%%%%%%%%%%% fig 6 %%%%%%%%%%%%%%%%%%%%%%%%%%%%%%%%%%%%%%%%%%%%%%%%%
%[hbtp]
%%%%%%%%%%%%%%%%%%%%%%%%%%%%%%%%%%%%%%%%%%%%%%%%%%%%%%%%%%%%%%%%%%%%%%%%%%%%%%%%%%%%%%%%%%%%%%%%%%

In a radial equilibrium situation, centrifugal forces given by $- v_\varphi^2/r$, equal the gravitational pull $F(r)$. Neglecting viscous 
forces and gradient pressure, we simply have,

\begin{equation}
-\frac{v_\varphi^2}{r} = F(r).
\label{centr}
\end{equation}

The total gravitational force $F(r)$ at a point $r$ can be expressed as,

\begin{equation}
F(r) = \int_0^r \rho(r') \frac{\partial \zeta}{\partial r'} dr',
\label{TotalF}
\end{equation}
where the function $\zeta = \zeta(r; a, e)$ can be obtained from the expression of the gravitational potential
for an exterior point. In that region (where $\rho(r) = 0$), Poisson's equation reads 
$\partial^2 \Psi / \partial r^2 = 0$. Therefore, we obtain:

\begin{equation}
\zeta(r; a, e) = \frac{2 \pi G}{e^2} (1 - e^2)^{1/2} \left[  \frac{a}{r} (r^2 - e^2 a^2)^{1/2} - 
\frac{r}{e} \sin^{-1} \left(\frac{e a}{r} \right)   \right].
\label{Zeta}
\end{equation}

From the above equation, we take the derivative $\partial \zeta / \partial r$, and we replace it in the equation 
for $F(r)$ (Eq. \ref{TotalF}).
Using the equilibrium Equation (\ref{centr}), we find that

\begin{equation}
v_\varphi^2(r) = 4 \pi G (1 - e^2)^{1/2} \int_0^r \frac{\rho(a) a^2 da}{(r^2 - e^2 a^2)^{1/2}}.
\label{Subtituting1}
\end{equation}

In Eq. (\ref{Subtituting1}), the integration variable $a$, describes an equipotential satisfying  the relation $a^2 = x^2 + y^2$, 
while $r^2 = x^2+y^2 + z^2$.
Taking now  the limit for a flat disk approximation, which means that the ratio between the minor and major radius tends to
zero, i.e., $b/a \rightarrow 0$, and the volume density $\rho(r)$ must be replaced by the surface density $\Sigma(r)$. We
obtain,

\begin{equation}
v_\varphi^2(r) =  G  \int_0^r \frac{\Sigma(a) a^2 da}{(r^2 - a^2)^{1/2}}.
\label{Subtituting}
\end{equation}

Using the relation $\partial \Psi/\partial r = - v_\varphi^2 / r$ we express the gravitational potential created by 
the disk, as;

\begin{equation}
\frac{\partial \Psi}{\partial r} = -\frac{G}{r} \int_0^r \frac{\Sigma(a) a^2}{(r^2 - a^2)^{1/2}} dr.
\label{PotFinal}
\end{equation}

This is the well know \textit{flat disk approximation}, for further reading about this 
approximation see Brandt \& Belton  (1962), and for mathematical details about the potential theory
O.D. Kellogg  (1956).

From a direct calculation, we can obtain the mass distribution of the disk $M_d(r)$ inside the radius $r$
by integration of the differential mass $dM(r) = 2 \pi r \Sigma(r) dr$, thus obtaining

\begin{equation}
M_d(r) = \int_0^r 2 \pi \Sigma(a) a da.
\label{MassInside}
\end{equation}

%%%%%%%%%%%%%%%%%%%%%%%%%%%%%%%%%%%%%%%%%%%%%%%%%

We calculate  the radial acceleration due the gravitational influence of the disk (i.e., $g_r = - \partial \Psi_{disk} / \partial r$)
using the Eq. \ref{PotFinal}. As for the vertical component of the gravitational acceleration 
produced by the disk, we have

\begin{equation}
g_z^{disk} = 2 \pi G \Sigma \frac{z}{H},
\label{gz}
\end{equation}
where $H$ is the effective scale of height of the disk.

The contribution from the black hole to the radial and vertical acceleration, $g_r^{bh}$ and $g_z^{bh}$ are, respectively,

\begin{equation}
g_r^{bh} = - \frac{G M_{bh}}{(r - r_s)^2} \hspace{1.5cm} \texttt{and} \hspace{1.5cm} g_z^{bh} = \frac{G M_{bh}}{r^3} z,
\label{gz}
\end{equation}
where $r_s = 2 G M_{bh}/c^2$ is the Schwarzschild radius.
The expression for $g_r^{bh}$ comes from the Paczy\'nski-Wiita Potential 
Paczy\'nski \& Wiita (1980). 

From these expressions, a simple parameter can be used to quantify the influence of the self-gravity.
We define two quantities by the ratios; 
$\varsigma_r \equiv g_r^{disk} / g_r^{bh}$, and $\varsigma_z \equiv g_z^{disk} /g_z^{bh} $.

If we consider an initial surface density profile given by a power law,

\begin{equation}
\Sigma(r) = \Sigma_0 \left( \frac{r}{r_0} \right)^{-n},
\label{DensityProf}
\end{equation}
the expression for  $\varsigma_z$ can be greatly simplified:

Firstly, we note that from the equation for the mass of the disk (\ref{MassInside}), we have

\begin{equation}
M_d(r) = \int_0^r 2 \pi \Sigma_0 \left( \frac{r'}{r_0} \right)^{-n} dr' = 
2 \pi \Sigma_0 \left( \frac{r}{r_0} \right)^{-n} \frac{r^2}{2 - n}.
\label{MassInside3}
\end{equation}

The expression for $\varsigma_z$ now reads,

\begin{equation}
\varsigma_z  = \frac{2 \pi \Sigma(r) r^3}{H M_{bh}} = 2 \pi \Sigma_0 \left( \frac{r}{r_0} \right)^{-n} r^2 
\left( \frac{r}{H} \right)  \frac{1}{M_{bh}},
\label{gzPar1}
\end{equation}
and using  Eq. (\ref{MassInside3}), we obtain

\begin{equation}
\varsigma_z  = (2 - n) ~ \frac{M_d(r)}{M_{bh}} ~ \frac{r}{H}.
\label{gzPar2}
\end{equation}

The above equation indicate that the disk becomes self-gravitating in the vertical direction when $M_d(r) / M_{bh}(r) \gtrsim H / r$.
Under normal conditions, the aspect ratio $H/r$ of the disk takes values in the range $ \sim 10^{-1} - 10^{-3}$, indicating that
even when the mass of the disk is smaller than the black hole mass, self gravity plays an important role and should be considered.

A similar condition is obtained in terms of the Toomre parameter $Q$ (Toomre  1964).
Toomre established the basic principles concerning the gravitational instabilities in accretion disks.
A gaseous disk is locally unstable to axisymmetric perturbations, if

\begin{equation}
Q = \frac{k c_s}{\pi G \Sigma} \lesssim 1,
\label{ToomreQ}
\end{equation}
where $k$ is the \textit{epicycle frequency}. This frequency  comes from small oscillations of a particle motion 
(infinitesimal perturbations) in a stable circular orbit without changing its angular momentum. 
The radial motion around the central object is periodic in that orbit,
called \textit{epicyclic oscillation}, his frequency is then called \textit{epicyclic frequency}. It is defined by:

\begin{equation}
k^2 = 2 \Omega^2 \left( 2 + \frac{d\ln \Omega}{d \ln r}  \right),
\label{EpiCyc}
\end{equation}
where the angular velocity $\Omega$ should be taken from the total gravitational potential $\Psi_t = \Psi_{bh} + \Psi_{disk}$, using
$\Omega^2 = \frac{1}{r} \frac{\partial \Psi_t}{\partial r} $. In the case of Keplerian rotation, we have $k = \Omega_K = \sqrt{G M_{bh}/r^3}$.

To see more clearly what lies behind the  $Q$ parameter, we take the following approximation: 
We assume that the disk approaches  a 
Keplerian rotational velocity (which, of course, is not realistic for a self-gravitating regimen), i.e., $\Omega \approx \Omega_K$. In 
this case, the epicycle frequency is $k \approx \Omega_K$ and  $c_s \approx H \Omega_K$. Now,
from the definition of the Toomre parameter $Q$, using the density profile given by (\ref{DensityProf}),
and the mass of the disk $M_d(r)$ as a function of $\Sigma(r)$ given in Eq. (\ref{MassInside3}) (assuming $n=1$), we have

\begin{eqnarray}
Q  & = & \frac{c_s \Omega_K}{\pi G \Sigma} \frac{r^2}{r^2} \nonumber\\
   & = & \frac{c_s \Omega_K}{\pi G \Sigma_0 \left( \frac{r}{r_0} \right)^{-n} r^2} \nonumber \\
   & = & 2 \left(  \frac{H}{r} \right)  \frac{M_{bh}(r)}{M_d(r)}(2 - n) \nonumber \\
   &  \cong & 2 \left( \frac{H}{r}\right)  \frac{M_{bh}(r)}{M_d(r)}. 
\label{ToomreBehind}
\end{eqnarray}

Again appears the ratio ${M_{bh}(r)}/{M_d(r)}$, indicating that the self-gravity begins to be important
starting from ${M_{d}(r)}/{M_{bh}(r)} \approx H/r$.
Recall that the above approximation was setting assuming Keplerian motion, therefore $k = \Omega_K$; so, what
Eq. (\ref{ToomreBehind}) indicate 
is a limit between a self-gravitating and a Keplerian regime for the scale of height. The Toomre parameter $Q$, defines
the gravitational \textit{instability threshold} under the condition $Q \gtrsim 1$ for stable regime.

\subsection{Other methods}

There are also other methods to solve Poisson's equation. For instance, development over the Green function, in which the term
$1 / (\mid \bf{x} - \bf{x'} \mid) $ in the Eq. (\ref{PoissonSol1}) is decomposed as,

\begin{equation}
\frac{1}{(\mid \bf{x} - \bf{x'} \mid)} = \sum_{l = 0}^\infty \sum_{m = -l}^l \frac{4 \pi}{2l + 1}
\frac{r^l_<}{r_>^{l+1}} Y^*_{lm}(\theta', \varphi) Y_{lm}(\theta, \varphi),
\label{Green1}
\end{equation}
where $r_< =\texttt{Inf}(r,r') $, $r_> =\texttt{Sup}(r,r') $ and $Y_{lm}(\theta, \varphi)$ are the  spherical harmonics used as
base, see for instance Cohl \& Tohline  (1999). 

A similar approach can be done using Bessel functions or 
using Fast Fourier Transform (FFT) methods (e.g. Binney \& Tremaine (1988), this last one is very popular
in 2D self-gravitating accretion disk models. The drawbacks in the FFT approach are linked to the precision (is a first order method), 
but this can be improved increasing grid number. This is not an issue for relatively small gaseous disk models (as protoplanetary disks),
but for larger disks (ranging from $\rm km$ to $\rm pc$) numerical cost using those methods are unavailable for most of the computer clusters.

%%%%%%%%%%%%%%%%%%%%%%%%%%%%%%%%%%%%%%%%%%%%%%%%referencias%%%%%%%%%%%%%%%%%%%%%%%%%%%%%%

%%%%%%%%%%%%%%%%%%%%%%%

\end{document}